
\input harvmac
\input epsf
\def\Re{{\rm Re}}
\def\Im{{\rm Im}}
\def\ra{\rightarrow}
\def\U{{\cal U}}
\def\L{{\cal L}}
\def\O{{\cal O}}
\def\gsim{{~\raise.15em\hbox{$>$}\kern-.85em
          \lower.35em\hbox{$\sim$}~}}
\def\lsim{{~\raise.15em\hbox{$<$}\kern-.85em
          \lower.35em\hbox{$\sim$}~}}
\def\epsK{\varepsilon_K}
\def\epe{\varepsilon^\prime/\varepsilon}
\def\tgl{{\tilde g}}
\def\dg{\Delta \Gamma}
\def\dm{\Delta M}
\def\gev{{\rm GeV}}
\def\tev{{\rm TeV}}
\def\gws{SU(3)\times SU(2) \times U(1)}
\def\H{{\cal H}}
\def\thc{{\theta_C}}

\def\YGTitle#1#2{\nopagenumbers\abstractfont\hsize=\hstitle\rightline{#1}%
\vskip .4in\centerline{\titlefont #2}\abstractfont\vskip .3in\pageno=0}
\YGTitle{SLAC-PUB-7379, WIS-96/49/Dec-PH, CERN-TH/96-368, hep-ph/9701231}
{\vbox{\centerline{CP Violation Beyond the Standard Model}}}
\smallskip
\centerline{Yuval Grossman$^a$,
Yosef Nir$^b$ and Riccardo Rattazzi$^c$}
\smallskip
\bigskip
\centerline{\it $^a$Stanford Linear Accelerator Center,
 Stanford University, Stanford, CA 94309, USA}
\centerline{\it $^b$Department of Particle Physics,
 Weizmann Institute of Science, Rehovot 76100, Israel}
\centerline{\it $^c$Theory Division, CERN, CH-1211 Geneva 23,
 Switzerland}
\bigskip
\bigskip
\baselineskip 18pt
\noindent
We review CP violation in various extensions of the electroweak sector
of the Standard Model. A particular emphasis is put on supersymmetric
models. We describe the two CP problems of supersymmetry, concerning
$d_N$ and $\epsK$. We critically review the various mechanisms that
have been suggested to solve these problems: exact universality,
approximate CP symmetry, alignment, approximate universality and
heavy squarks. We explain how future measurements of CP violation will
test these mechanisms. We describe extensions of the quark sector and
their implications on CP asymmetries in neutral $B$ decays, on the
$K_L\to\pi\nu\bar\nu$ decay and on $\Delta\Gamma(B_s)$. We discuss
CP violation in charged scalar exchange in models with natural flavor
conservation and explain how transverse lepton polarization in meson
decays can probe such models. CP violation in neutral scalar exchange
arises in models of horizontal symmetries
and may be manifest in heavy quark ($b$ and $t$) physics.
We describe the implications of Left-Right Symmetric models on
$d_N$, $\epsK$, $\epe$ and CP asymmetries in $B$ decays. Finally,
we briefly discuss the potential of future measurements of CP violation
to discover New Physics.
\bigskip
\bigskip
\baselineskip 15pt
\leftskip=1.2cm\rightskip=1.2cm\noindent {\bf
To appear in the Review Volume ``Heavy Flavours II'',
eds. A.J. Buras and M. Lindner,
Advanced Series on Directions in High Energy Physics,
World Scientific Publishing Co., Singapore.}

\baselineskip 18pt
\leftskip=0cm\rightskip=0cm

\Date{}

\newsec{Introduction}

CP violation is one of the most promising directions in the search for
New Physics beyond the Standard Model.

$\bullet$ Experimentally, the Standard Model (SM) picture of CP violation
has not been tested yet. At present, there is only a single (complex)
CP violating parameter that has been measured
\ref\CCFT{J.H. Christenson, J.W. Cronin, V.L. Fitch and R. Turlay,
 Phys. Rev. Lett. 13 (1964) 138.}.
This is $\epsK$ of the neutral $K$ system. Within the Standard Model,
the existing measurements merely fix the value of the CP violating phase
$\delta_{\rm KM}$ in the Cabibbo-Kobayashi-Maskawa (CKM) mixing matrix
for quarks
\ref\KoMa{M. Kobayashi and T. Maskawa, Prog. Theo. Phys. 49 (1973) 652.}\
but cannot test the prediction that $\delta_{\rm KM}$ constitutes the
{\it only} source of  CP violation.\foot{A large value of $\epsK$
would be inconsistent with the SM. However, any value
$|\epsK|\leq\O(10^{-3})$ can be accommodated.} A genuine testing
of the KM picture of CP violation awaits the building of $B$ factories
that would provide a second, independent measurement of CP violation
\ref\BCP{A.B. Carter and A.I. Sanda, Phys. Rev. Lett. 45 (1980) 952;
 Phys. Rev. D23 (1981) 1567.}.

$\bullet$ The observed baryon asymmetry of the Universe, if dynamically
generated, requires that CP is violated
\ref\Sakh{A.D. Sakharov, ZhETF Pis. Red. 5 (1967) 32;
 JETP Lett. 5 (1967) 24.}.
The Standard Model CP violation, closely related to highly suppressed
flavor changing processes, fails to produce this asymmetry by
many orders of magnitude. (For recent discussions, see
\nref\FS{G.R. Farrar and M.E. Shaposhnikov, Phys. Rev. D50 (1994) 774.}%
\nref\Gave{M.B. Gavela {\it et al.}, Nucl. Phys. B430 (1994) 382.}%
\nref\HuSa{P. Huet and E. Sather, Phys. Rev. D51 (1995) 379.}%
\refs{\FS-\HuSa}.) In contrast, various extensions of the SM and,
in particular, models of New Physics close to the electroweak scale,
provide new sources of CP violation that are large enough to be
consistent with the observed asymmetry (for a recent review, see
\ref\CKN{A.G. Cohen, D.B. Kaplan and
A.E. Nelson,  Ann. Rev. Nucl. Part. Sci. 43 (1993) 27.}).
In some models, phases that are large enough to generate the
baryon asymmetry also induce an electric dipole moment (EDM) of the
neutron not far below the present experimental bound
\nref\EGNR{J. Ellis, M.K. Gaillard, D.V. Nanopoulos and S. Rudaz,
 Phys. Lett. 99B (1981) 101.}%
\nref\HuNe{P. Huet and A.E. Nelson, Phys. Lett. B355 (1995) 229.}%
\refs{\EGNR,\HuNe}.

$\bullet$ The QCD lagrangian does allow an additional source of
CP violation, that is the $\theta_{\rm QCD}$ parameter. However, an
extreme fine-tuning is needed in order that its contribution to
the electric dipole moment of the neutron $d_N$ does not exceed
the experimental upper bound. Various mechanisms that go beyond
the SM, e.g. a Peccei-Quinn symmetry, spontaneous CP violation,
or a vanishing mass of the up quark, may solve the problem.

$\bullet$ Almost any extension of the SM has, in general, new
CP violating phases.

In this review we describe various extensions of the Standard Model
and their implications on CP violation. We do not discuss in any detail
either baryogenesis or the strong CP problem. Instead, we focus on
extensions of the electroweak sector and their implications for
the EDM of the neutron which, if measured in the foreseeable future, will
clearly signal New Physics, and for CP violation in neutral meson mixing
which, in some cases, is free from hadronic uncertainties and therefore
could distinguish between Standard Model and New Physics contributions.

In section 2 we give a detailed and critical discussion of CP violation
(and flavor problems) in Supersymmetry. We describe the supersymmetric CP
problem, that is the generically too large supersymmetric contributions
to the electric dipole moment of the neutron, and the supersymmetric
$\epsK$ problem. We present five classes of supersymmetric models
that solve or relax these problems: exact universality, approximate
CP symmetry, alignment, approximate universality and heavy squarks.
We explain how future measurements of CP violation will test these
models.

Section 3 is devoted to extensions of the fermion sector. In particular,
we consider additional $SU(2)$-singlet down quarks. Our emphasis
here is on CP asymmetries in neutral $B$ decays into final CP
eigenstates. These can be dramatically modified in such extensions.
For certain decay modes, the measurement of the asymmetries can cleanly
determine the relevant parameters of the extended sector.
We also discuss the decay $K_L\to\pi\nu\bar\nu$ and CP violation
in the width difference $\Delta\Gamma (B_s)$.

In section 4 we discuss extensions of the scalar sector. A model of
spontaneous CP violation and Natural Flavor Conservation (NFC), where
CP violation arises from charged scalar exchange only,
provides us with an example of how CP violation
can actually rule out various extensions of the Standard Model.
We also explain how CP violation in charged scalar exchange may
affect transverse lepton polarization.
Then we discuss CP violation in neutral scalar exchange in models
where approximate flavor symmetries (invoked to explain the
smallness of the quark and lepton masses and mixing) replace
NFC in suppressing flavor changing neutral current (FCNC) processes.
We finally describe the idea of superweak CP violation with emphasis
on the fact that it refers to many different types of models.

In section 5 we briefly discuss an extension of the gauge sector.
We describe a Left-Right Symmetric model (LRS) where CP is
spontaneously broken. We argue that, when CP violation arises
in non-horizontal gauge interactions, the effects in the $B$ system
are likely to be small.

Our conclusions are given in section 6, where we present the various
future measurements of CP violation with emphasis on their potential
to discover effects of New Physics.

\newsec{Supersymmetry}

A generic supersymmetric extension of the Standard Model contains a host
of new flavor and CP violating parameters
(for reviews on supersymmetry see
\nref\nilles{H.P. Nilles, Phys. Rep. 110 (1984) 1.}%
\nref\haberkane{H.E. Haber and G.L. Kane, Phys. Rep. 117 (1985) 75.}%
\nref\barbieri{R. Barbieri, Riv. Nuovo Cim. 11 (1988) 1.}%
\nref\HaberTASI{H.E. Haber, SCIPP 92/33, Lectures given at TASI 92.}%
refs. \refs{\nilles-\HaberTASI}). The requirement of consistency
with experimental data provides strong constraints on many of these
parameters. For this reason, the physics of flavor and CP violation
has had a profound impact on supersymmetric model building. A discussion
of CP violation in this context can hardly avoid addressing the flavor
problem itself.  Indeed, many of the supersymmetric models that we
analyze below were originally aimed at solving flavor problems.

As concerns CP violation, one can distinguish two classes of experimental
constraints. First, bounds on nuclear and atomic electric dipole moments
determine what is usually called the {\it supersymmetric CP problem}.
Second, the physics of neutral mesons and, most importantly, the small
experimental value of $\epsK$ pose the {\it supersymmetric $\epsK$
problem}. In the next two subsections we describe the two problems.

In most of the literature, solutions to these two problems are discussed
separately. We believe, however, that since they represent the same
issue, {\it i.e.} the origin of CP violation, and since, in general, the
mechanisms that solve them can be classified in similar ways, they should
be discussed together to get the appropriate picture of SUSY CP
violation. Thus, we analyze in turn five classes of supersymmetric models
and all aspects of CP violation for each of them.

Before turning to a detailed discussion, we define two scales that
play an important role in supersymmetry: $\Lambda_S$, where
the soft supersymmetry breaking terms are generated, and $\Lambda_F$,
where flavor dynamics takes place. When $\Lambda_F\gg\Lambda_S$, it is
possible that there are no genuinely new sources of flavor and CP
violation. This leads to models with exact universality, which we
discuss in section 2.3. When $\Lambda_F\lsim\Lambda_S$, we do not
expect, in general, that flavor and CP violation are limited to the
Yukawa matrices. One way to suppress CP violation would be to assume
that CP is an approximate symmetry of the full theory (namely, CP
violating phases are all small). We discuss this scenario in section 2.4.
Another option is to assume that, similarly to the Standard Model,
CP violating phases are large, but their effects are screened, possibly
by the same physics that explains the various flavor puzzles.
Such models, with Abelian or non-Abelian horizontal symmetries,
are described in sections 2.5 and 2.6, respectively. Finally, it is
possible that CP violating effects are suppressed because squarks
are heavy. This scenario is discussed in section 2.7.

\subsec{The Supersymmetric CP Problem}

One aspect of supersymmetric CP violation involves effects that are
flavor preserving. Then, for simplicity, we describe this aspect in
a supersymmetric model without additional flavor mixings, {\it i.e.} the
minimal supersymmetric standard model (MSSM) with universal sfermion
masses and with the trilinear SUSY-breaking scalar couplings
proportional to the corresponding Yukawa couplings. (The generalization
to the case of non-universal soft terms is straightforward.)  In such a
constrained framework, there are four new phases beyond the two
phases of the Standard Model ($\delta_{\rm KM}$ and $\theta_{\rm QCD}$).
One arises in the bilinear $\mu$-term of the superpotential,
\eqn\muterm{W=\mu H_uH_d,}
while the other three arise in the soft supersymmetry breaking
parameters $m_\tgl$ (the gaugino mass), $A$ (the trilinear
scalar coupling) and $m_{12}^2$ (the bilinear scalar coupling):
\eqn\sSUSYb{\L=-{1\over2}m_\tgl\tgl\tgl-A(Y^u QH_u\bar u
-Y^d QH_d\bar d-Y^e LH_d\bar\ell)-m_{12}^2 H_uH_d+{\rm h.c.},}
where $\tgl$ are the gauginos and $Y$ are Yukawa matrices.
Only two combinations of the four phases are physical
\ref\DGH{M. Dugan, B. Grinstein and L.J. Hall,
 Nucl. Phys. B255 (1985) 413.}.
This can be easily shown by following the discussion of ref.
\ref\DiTh{S. Dimopoulos ans S. Thomas, Nucl. Phys. B465 (1996) 23.}.
In the absence of \muterm\ and \sSUSYb, there are two additional global
$U(1)$ symmetries in the MSSM, an $R$ symmetry and a Peccei-Quinn
symmetry. This means that one could treat the various dimensionful
parameters in \muterm\ and \sSUSYb\ as spurions which break the
symmetries, thus deriving selection rules. The appropriate
charge assignments are:
\eqn\spur{\matrix{
&m_\tgl&A&m_{12}^2&\mu&H_u&H_d&Q\bar u&Q\bar d&L\bar\ell\cr
U(1)_{\rm PQ}&0&0&-2&-2&1&1&-1&-1&-1\cr
U(1)_{\rm R}&-2&-2&-2&0&1&1&1&1&1\cr}}
Physical observables can only depend on combinations of the dimensionful
parameters that are neutral under both $U(1)$'s. There are three such
independent combinations: $m_\tgl\mu(m_{12}^2)^*$, $A\mu(m_{12}^2)^*$
and $A^* m_\tgl$. However, only two of their phases are independent, say
\eqn\phiAB{\phi_A=\arg(A^* m_\tgl),\ \ \
\phi_B=\arg(m_\tgl\mu(m_{12}^2)^*).}
In the more general case of non-universal soft terms there is
one independent phase $\phi_{A_{i}}$ for each quark and lepton flavor.
Moreover, complex off-diagonal entries in the sfermion
mass matrices may represent additional sources of CP violation.

The most significant effect of $\phi_A$ and $\phi_B$ is their
contribution to electric dipole moments (EDMs). For example, the
contribution from one-loop gluino diagrams to the down quark EDM
is given by
\nref\BuWy{W. Buchmuller and D. Wyler, Phys. Lett. B121 (1983) 321.}%
\nref\PoWi{J. Polchinski and M. Wise, Phys. Lett. B125 (1983) 393.}%
\refs{\BuWy-\PoWi}:
\eqn\ddsusy{d_d=M_d{e\alpha_3\over 18\pi\tilde m^4}\left(
|A m_{\tilde g}|\sin\phi_A+\tan\beta|\mu m_{\tilde g}|\sin\phi_B\right),}
where we have taken $m^2_Q\sim m^2_D\sim m^2_{\tilde g}\sim\tilde m^2$,
for left- and right-handed squark and gluino masses. We define, as usual,
$\tan\beta = \vev{H_u}/\vev{H_d}$. Similar one-loop diagrams give rise to
chromoelectric dipole moments. The electric
and chromoelectric dipole moments of the light quarks $(u,d,s)$ are the
main source of $d_N$ (the EDM of the neutron), giving
\ref\FPT{W. Fischler, S. Paban and S. Thomas,
 Phys. Lett. B289 (1992) 373.}\
\eqn\dipole{d_N\sim 2\, \left({100\, \gev\over \tilde m}\right )^2
\sin \phi_{A,B}\times10^{-23}\ e\, {\rm cm}}
where, as above, $\tilde m$ represents the overall SUSY scale.
The present experimental bound, $d_N<1.1\times 10^{-25}e\, {\rm cm}$
\nref\smith{K.F. Smith {\it et al.}, Phys. Lett. B234 (1990) 191.}%
\nref\altarev{I.S. Altarev {\it et al.}, Phys. Lett. B276 (1992) 242.}%
\refs{\smith,\altarev}, is then violated for  $\O(1)$ phases, unless the
masses of superpartners are above $\O(1\,\tev)$. Alternatively for light
SUSY masses, the new phases should be $<\O(10^{-2})$. Notice however that
one may consider the actual bound weaker than this, due to the
theoretical uncertainty in the estimate of the hadronic matrix elements
that lead to eq. \dipole\ (see ref.
\ref\florellis{J. Ellis and R. Flores, Phys. Lett. B377 (1996) 83.}\
for a recent discussion of possible cancellations among the contributions
to $d_N$ in the case of a universal $\phi_A$). With this caveat, whether
the phases are small or squarks are heavy, a fine-tuning of order
$10^{-2}$ seems to be required, in general, to avoid too large a $d_N$.
This is {\it the Supersymmetric CP Problem}
\nref\EFN{J. Ellis, S. Ferrara and D. Nanopoulos,
 Phys. Lett. B114 (1982) 231.}%
\nref\Barr{S. Barr, Int. J. Mod. Phys. A8 (1993) 209.}%
\refs{\BuWy-\PoWi,\EFN-\Barr}.

In addition to $d_N$, the SUSY CP phases contribute to atomic and nuclear
EDMs (see a detailed discussion in ref. \FPT).
The former are also sensitive to phases in the leptonic sector.
The latter give additional constraints on the quark sector phases.
For instance, the bound on the nuclear EDM of ${}^{199} {\rm Hg}$
is comparable to the one given by $d_N$.
 In practice, these additional bounds on SUSY CP phases
are not stronger than those from $d_N$ (at least in the quark sector).
However, since there are significant theoretical uncertainties in the
calculation of nuclear EDMs, it is important to measure as many as
possible of them to obtain more reliable bounds.

\subsec{The Supersymmetric $\epsK$ Problem}

The contribution to the CP violating $\epsK$ parameter in the neutral $K$
system is dominated by diagrams involving $Q$ and $\bar d$ squarks in the
same loop
\nref\DNW{J. Donoghue, H. Nilles and D. Wyler,
 Phys. Lett. B128 (1983) 55.}%
\nref\GaMa{F. Gabbiani and A. Masiero, Nucl. Phys. B322 (1989) 235.}%
\nref\HKT{J.S. Hagelin, S. Kelley and T. Tanaka,
 Nucl. Phys. B415 (1994) 293.}%
\nref\GMS{E. Gabrielli, A. Masiero and L. Silvestrini,
 Phys. Lett. B374 (1996) 80.}%
\nref\GGMS{F. Gabbiani, E. Gabrielli, A. Masiero and L. Silvestrini,
Nucl. Phys. B477 (1996) 321.}%
\refs{\DNW-\GGMS}. The corresponding effective four-fermi operator
involves fermions of both chiralities, so that its matrix elements
are enhanced by $\O(m_K/m_s)^2$ compared to the chirality conserving
operators. For $m_{\tilde g}\simeq m_Q \simeq m_D= \tilde m$ (our results
depend only weakly on this assumption) and focusing on the contribution
from the first two squark families, one gets (we use the results in
ref. \GGMS)
\eqn\epsKSusy{\epsK={5\ \alpha_3^2  \over 162\sqrt2}{f_K^2m_K\over\tilde
m^2\Delta m_K}\left [\left({m_K\over m_s+m_d}\right)^2+{3\over 25}\right]
\Im\left\{{(\delta m_Q^2)_{12}\over m_Q^2}
{(\delta m_D^2)_{12}\over m_D^2}\right\},}
where $(\delta m_{Q,D}^2)_{12}$ are the off diagonal entries in the
squark mass matrices in a basis where the down quark mass matrix
and the gluino couplings are diagonal. These flavor
violating quantities are often written as
$(\delta m_{Q,D}^2)_{12}=V_{11}^{Q,D}\delta m_{Q,D}^2 V_{21}^{Q,D*}$,
where $\delta m_{Q,D}^2$ is the mass splitting among the squarks
and $V^{Q,D}$ are the gluino coupling mixing matrices in the mass
eigenbasis of quarks and squarks. Note that CP would be violated
even if there were two families only
\ref\NirSusy{Y. Nir, Nucl. Phys. B273 (1986) 567.}.
There are also contributions involving the third family squarks via the
(13) and (23) mixings. In some cases the third family contribution
actually dominates. Using the experimental value of $\epsK$,
we get the constraint
\eqn\epsKScon{\left ({300 \ \gev\over\tilde m}\right)^2\left|
{(\delta m_Q^2)_{12} \over m_Q^2}{(\delta m_D^2)_{12}
\over m_D^2}\right|\sin\phi\lsim 0.5\times 10^{-7},}
where $\phi=\arg((\delta m_Q^2)_{12} (\delta m_D^2)_{12} )$.
In a generic supersymmetric framework, we expect $\tilde m=\O(m_Z)$,
$\delta m_{Q,D}^2/m_{Q,D}^2=\O(1)$, $V_{ij}^{Q,D}=\O(1)$ and
$\sin\phi=\O(1)$. Then the constraint \epsKScon\ is generically violated
by about seven orders of magnitude. Four-fermi operators with same
chirality fermions give a smaller effect.
The resulting $\epsK$-bounds are therefore somewhat weaker:
\eqn\samechir{ \left ({300 \ \gev\over m_Q}\right)^2\left |\Im\left(
{(\delta m_Q^2)_{12}\over m_Q^2}\right)^2\right|\lsim 10^{-6}}
and similarly for the right handed squarks.

\subsec{Exact Universality}

Both supersymmetric CP problems are solved if, at the scale $\Lambda_S$,
the soft supersymmetry breaking terms are universal and the genuine SUSY
CP phases $\phi_{A,B}$ vanish. Then the Yukawa matrices represent
the only source of flavor and CP violation which is relevant in low
energy physics. This situation can naturally arise when supersymmetry
breaking is mediated by gauge interactions at a scale
$\Lambda_S\ll\Lambda_F$
\nref\DiNe{M. Dine and A. Nelson, Phys. Rev. D48 (1993) 1277.}%
\nref\DNeS{M. Dine and A. Nelson and Y. Shirman,
 Phys. Rev. D51 (1994) 1362.}%
\nref\DNNS{M. Dine, A. Nelson, Y. Nir and Y. Shirman,
 Phys. Rev. D53 (1996) 2658.}%
 \nref\DNiS{M. Dine, Y. Nir and Y. Shirman,
 Phys. Rev. D55 (1997) 1501.}%
\refs{\DiNe-\DNiS}. In the simplest scenarios, the $A$-terms and the
gaugino masses are generated by the same SUSY and $U(1)_R$ breaking
source (see eq. \spur). Thus, up to very small effects due to the
{\it standard}  Yukawa matrices, $\arg(A)=\arg(m_{\tilde g})$ so that
$\phi_A$ vanishes. In specific models also $\phi_B$ vanishes in a similar
way \refs{\DNeS,\DNiS}. It is also possible that similar boundary
conditions occur when supersymmetry breaking is communicated to the
observable sector up at the Planck scale
\nref\CAN{A.H. Chamseddine, R. Arnowitt and P. Nath,
 Phys. Rev. Lett. 49 (1982) 970; Nucl. Phys. B227 (1983) 1219.}%
\nref\BFS{R. Barbieri, S. Ferrara and C.A. Savoy,
 Phys. Lett. B119 (1982) 343.}%
\nref\HLW{L. Hall, J. Lykken and S. Weinberg,
 Phys. Rev. D27 (1983) 235.}%
\nref\EKN{J. Ellis, C. Kounnas and D.V. Nanopoulos, Nucl. Phys. B247
 (1984) 373.}%
\nref\LaRo{M. Lanzagorta and G.G. Ross, Phys. Lett. B364 (1995) 163.}%
\refs{\CAN-\LaRo}. The situation in this case seems to be less under
control from the theoretical point of view. Dilaton dominance in SUSY
breaking, though, seems a very interesting direction to explore
\nref\KaLo{V. Kaplunovsky and J. Louis, Phys. Lett. B306 (1993) 269}%
\nref\BML{R. Barbieri, J. Louis and M. Moretti,
 Phys. Lett. B312 (1993) 451; (E) {\it ibid.} B316 (1993) 632.}%
\refs{\KaLo,\BML}.

The most important implication of this type of boundary conditions
for soft terms, which we refer to as {\it exact  universality}
\nref\DiGe{S. Dimopoulos and H. Georgi, Nucl. Phys. B193 (1981) 150.}%
\nref\Saka{N. Sakai, Z. Phys. C11 (1981) 153.}%
\refs{\DiGe-\Saka},
is the existence of the SUSY analogue of the GIM mechanism which operates
in the SM. The CP violating phase of the CKM matrix can
feed into the soft terms via Renormalization Group (RG) evolution only
with a strong suppression from light quark masses \DGH.

With regard to the supersymmetric CP problem, gluino diagrams
contribute to quark EDMs as in eq. \ddsusy,
but with a highly suppressed effective phase, {\it e.g.}
\eqn\gim{\phi_{A_d}\sim (t_S/16 \pi^2)^4 Y_t^4 Y_c^2 Y_b^2 J.}
Here $t_S=\log (\Lambda_S/M_W)$ arises from the RG evolution from
$\Lambda_S$ to the electroweak scale, the $Y_i$'s are quark Yukawa
couplings (in the mass basis), and
$J={\rm Im}(V_{ud}V_{tb}V_{ub}^*V_{td}^*)\simeq 2\times 10^{-5}$
is the invariant measure of CP violation in the CKM matrix
\ref\Jarl{C. Jarlskog, Phys. Rev. Lett. 55 (1985) 1039.}.
A similar contribution comes from chargino diagrams. The resulting EDM is
$d_N\lsim 10^{-31}\ e\ {\rm cm}$. This maximum can be reached only for
very large $\tan\beta\sim60$ while, for small $\tan\beta \sim 1$,
$d_N$ is about 5 orders of magnitude smaller. This range of values for
$d_N$ is much below the present ($\sim10^{-25}\ e$ cm) and foreseen
($\sim 10^{-28}\ e$ cm) experimental sensitivities.  The smallness
of these contributions has been recently emphasized in ref.
\ref\RoStru{A. Romanino and A. Strumia, hep-ph/9610485.},
\nref\BeVi{S. Bertolini and F. Vissani, Phys. Lett. B324 (1994) 164.}%
\nref\IMSS{T. Inui, Y. Mimura, N. Sakai and T. Sasaki,
 Nucl. Phys. B449 (1995) 49.}%
\nref\ACW{S.A. Abel, W.N. Cottingham and I.B. Wittingham,
 Phys. Lett. B370 (1996) 106.}%
by using the spurionic analysis of ref. \DGH\ which keeps the GIM
mechanism manifest. (For previous numerical estimates of the effective
phases, see refs. \refs{\BeVi-\ACW}.)

With regard to the supersymmetric $\epsK$ problem, the contribution
to $\epsK$ is proportional to $\Im(V_{td}V_{ts}^*)^2 Y_t^4
(t_S/16 \pi^2)^2$, giving the same GIM suppression as in the SM.
This contribution turns out to be small. Using the result in Ref. \DGH,
we get
\eqn\unieps{|\epsK^{\rm SUSY}|\sim 6 \times 10^{-6} \left [{J\,\Re
(V_{td}V_{ts}^*)\over10^{-8}}\right]\left[{300\gev\over\tilde m}\right]^2
\left [ {\ln (\Lambda_S/m_W)\over 5}\right ]^2.}
The value $t_S=5$ is typical to gauge mediated supersymmetry breaking,
but \unieps\ remains negligible for any scale $\Lambda_S\lsim
M_{\rm Pl}$ (namely $t_S\lsim 35$). The supersymmetric contribution to
$D-\bar D$ mixing is similarly small and we expect no observable effects.

For the $B_d$ and $B_s$ systems, the largest SUSY contribution
to the mixing comes from box diagrams with intermediate charged Higgs and
the up quarks. It can be up to $\O(0.2)$ of the SM amplitude for
$\Lambda_S=M_{\rm Pl}$ and $\tan\beta = \O(1)$
\ref\BBMR{S. Bertolini, F. Borzumati, A. Masiero and G. Ridolfi,
Nucl. Phys. B353 (1991) 591.},
and much smaller for large $\tan\beta$. The contribution is smaller in
models of gauge mediated SUSY breaking where the mass of the charged
Higgs boson is typically $\gsim 300\ \gev$ \DNNS\ and $t_S\sim5$. The
SUSY contributions to $B_s\bar B_s$ and $B_d \bar B_d$ mixing are, to a
good approximation, proportional to $(V_{tb}V_{ts}^*)^2$ and
$(V_{tb}V_{td}^*)^2$, respectively, like in the SM. Then, regardless of
the size of these contributions, the relation
$\Delta m_{B_d}/\Delta m_{B_s}\sim |V_{td}/V_{ts}|^2$ and the
CP asymmetries in neutral
$B$ decays into final CP eigenstates are the same as in the SM.

\subsec{The Non-Universal Case: Approximate CP Symmetry}

Both supersymmetric CP problems are solved if CP is an approximate
symmetry, broken by a small parameter of order $10^{-3}$. This is one
of the possible solutions to CP problems in the class of supersymmetric
models with $\Lambda_F\lsim\Lambda_S$, where the soft masses are
generically not universal, so that we do not expect flavor and CP
violation to be limited to the Yukawa matrices.\foot{Of course,
some mechanism has also to suppress the real part of the $\Delta S=2$
amplitude by a sufficient amount.} Most models where soft terms arise at
the Planck scale ($\Lambda_S\sim M_{\rm Pl}$) belong to this class.

In order to have a successful mechanism to screen CP violating
phases, a theory or a set of assumptions on the origin of CP violation is
needed. Such a theory  has to be able to reproduce the
only well established evidence of CP violation in experimental data,
$\epsK$, without affecting in an unacceptable way all the other CP odd
observables. On this point, supersymmetric models (as many other
extensions of the standard model) provide us with two radically
different possibilities.
A first, perhaps {\it reactionary}, (as dubbed in ref.
\ref\HaWeS{L.J. Hall and S. Weinberg, Phys. Rev. D48 (1993) R979.}\
in the context of multi-Higgs models) point of view is  that the CKM
picture of CP violation is incorrect, {\it i.e.} that $\epsK\ll 1$ {\it
not} because of the smallness of mixing angles and quark mass differences
(GIM), but just because CP odd phases happen to be small wherever they
appear. In other words, CP is an approximate symmetry of the full theory,
not just of the standard model sector. (The second point of view is
described in the next two sections.)

If CP is an approximate symmetry, we expect also the SM
phase $\delta_{\rm KM}$ to be $\ll 1$. Then the standard box diagrams
cannot account for $\epsK$ which should arise from another
source. In supersymmetry with non-universal soft terms, the source could
be diagrams involving virtual superpartners, mainly squark-gluino box
diagrams. Let us call $(M_{12}^K)^{\rm SUSY}$
the supersymmetric contribution to the $K-\bar K$ mixing amplitude.
Then the requirements $\Re (M_{12}^K)^{\rm SUSY}\lsim\Delta m_K$
and $\Im(M_{12}^K)^{\rm SUSY}\sim\epsK\Delta m_K$ imply that the
generic CP phases are $\geq\O(\epsK)\sim 10^{-3}$.

Of course, $d_N$ constrains the relevant CP violating phases to be
$\lsim10^{-2}$. If all phases are of the same order, then $d_N$ must be
just below or barely compatible with the present experimental bound.
A signal should definitely be found if the accuracy is increased by two
orders of magnitude.

The main phenomenological implication of these scenarios is that
CP asymmetries in $B$ meson decays are small, perhaps
$\O(\epsK)$, rather than ${\cal O}(1)$ as expected in the SM.

Large deviations from the SM are also possible in $\epe$. Indeed, as can
be inferred from ref. \GGMS, when CP violation appears mainly in the
diagonal blocks of the squark mass-squared matrices, the constraint from
$\epsK$ implies $\epe\lsim 10^{-5}$. When there is considerable
CP violation also in $A$-terms or gaugino masses $\epe$ can be larger.

Ref.
\nref\abefre{S.A. Abel and J.M. Frere, Phys. Rev. D55 (1997) 1623.}%
\ref\BaBa{K.S. Babu and S.M. Barr, Phys. Rev. Lett. 72 (1994) 2831.}\
presents an interesting attempt to naturally generate an approximate CP
symmetry: CP is spontaneously broken in a sector of heavy fermions in
vector representations of $\gws$, and is transmitted to the MSSM only via
radiative corrections. The only resulting observable phases then appear
in  gaugino masses and are of order $\alpha_3/4\pi$ for gluinos and
$\alpha_2/4\pi$ for winos. So all CP violating effects are suppressed by
$\alpha/4\pi$, which seems very promising.\foot{See also ref. \abefre\
which discusses a particular ansatz where CP violation appears only in
$A$-terms.} However, in order to reproduce the correct  value of $\epsK$,
this model needs rather large $A$-terms, $A_{12}^d\gsim4Y_s\tilde m$ (the
naive expectation in most flavor models would be $A_{12}^d\sim\thc Y_s
\tilde m$ where $\thc\sim0.2$ is the Cabibbo angle). The resulting
contribution to EDMs depends on the flavor structure of the $A$-terms and
could be large even for small $\phi_A$. Such a large $A_{12}^d$ also
leads to $\epe\gsim3\times10^{-3}$, barely compatible with present
bounds.

\subsec{The Non-Universal Case: Approximate Abelian Horizontal
Symmetries}

For supersymmetric models with $\Lambda_F\lsim\Lambda_S$, where
there are genuine supersymmetric sources of flavor and CP violation,
one can still take a point of view that is very different from the one
described in the previous section: The CKM picture of CP violation is the
correct one, whereby $\epsK\sim 10^{-3}$ results from small flavor
mixings rather than small phases in the individual Lagrangian parameters.
We now expect ${\cal O}(1)$ phases, so that an explanation is needed for
the absence or smallness of the new supersymmetric contribution to
$\epsK$ and to the EDMs. Therefore, mechanisms to suppress FCNC and
to screen the CP phases in the soft SUSY breaking mass terms and
in the $A$ terms are required. Abelian horizontal symmetries, which
are invoked to explain the flavor structure of the observed quarks,
can provide at the same time CP screening mechanisms that are efficient
enough to solve both CP problems.

With regard to the neutral $K$ system, a possible mechanism to screen the
CP violating phases in the supersymmetric box diagrams is provided by
{\it alignment}
\ref\QSA{Y. Nir and N. Seiberg, Phys. Lett. B309 (1993) 337.}:
The squark mass matrices have a structure, but they have a reason to be
diagonal in the basis set by the quark mass matrix. This is achieved in
\nref\LNSb{M. Leurer, Y. Nir and N. Seiberg,
 Nucl. Phys. B420 (1994) 468.}%
models of Abelian horizontal symmetries \refs{\QSA,\LNSb}.
The symmetry is spontaneously broken by the VEVs of scalar fields
$\{\Phi\}$ (``flavons"), producing a small parameter $\lambda\equiv
\Phi/\Lambda_F$ which is usually taken to have a value $\lambda\sim0.2$.
This small parameter is responsible for the smallness and hierarchy of
quark masses and mixings. The solution of the supersymmetric $\epsK$
problem in this framework makes use of the fact that in
supersymmetric theories, the Yukawa matrices $Y^q$ must be holomorphic
functions of the flavon fields $\{\Phi\}$. By assigning appropriate
$\H$-charges to the quark superfields, holomorphy dictates that the
$2\times2$ down Yukawa sub-matrix is diagonal in the flavor basis,
{\it i.e.} the basis where the quark fields have definite $\H$ charges.
Then the only supersymmetric contributions to $K-\bar K$ mixing
arise directly via the (12) entries in $m_{Q,D}^2$ and, indirectly,
via the third family. The left hand side of eq. \epsKScon\ is of order
$\lambda^{12}\sim10^{-8}$, consistent with the bound.

A solution which does not require much universality is also
provided by the dynamical mechanism suggested in ref.
\ref\DGT{S. Dimopoulos, G.F. Giudice and N. Tetradis,
 Nucl. Phys. B454 (1995) 59.}\
(see also Ref.
\ref\SaRa{R. Rattazzi and U. Sarid, Nucl. Phys. B475 (1996) 27.}\
for a more critical discussion). There the soft terms correspond to
fields that are free to have different orientations in flavor space.
Their expectation value is dynamically determined by the only source of
``explicit'' flavor violation which is assumed to be relevant, the Yukawa
matrices. Then all SUSY induced CP violating effects end up being
proportional to $J\sim 2 \times 10^{-5}$. Moreover these effects appear
without additional suppression only in operators that do not involve
right handed fermions. Thus the leading effect has to satisfy the weaker
bound of eq. \samechir\ which, with the suppression from $J$, requires
only a mild squark  degeneracy.

An extension of these ideas, aimed at screening the CP phases in the
$A$-terms, is given in Ref.
\ref\RaNi{Y. Nir and R. Rattazzi, Phys. Lett. B382 (1996) 363.}.
Since we do not have universality, the phases $\phi_A$ are in general
different for each quark flavor:
\eqn\ephase{\phi_{A_q}=\arg\left({A_qm_{\tilde g}^*\over Y_q}\right).}
(Here, in the definition of $A$, we do not factor out the Yukawa matrices
$Y$, as was done instead in the universal case \sSUSYb.) In the model of
ref. \RaNi, CP is assumed to be a symmetry of the Lagrangian.
The flavon fields $\Phi$ spontaneously break not only $\H$ but also CP.
This assumption implies, in particular, that the soft terms, before the
breakdown of $\H$, can be all made real. Both the Yukawa matrices and
$A$-term matrices are flavon dependent and could be complex through
their dependence on $\Phi$. On the other hand, $m_{\tilde g}$ is a
flavor singlet and therefore real to a very good approximation. A crucial
property of supersymmetric theories is that both $A_q$ and $Y_q$ must be
holomorphic functions of the flavon fields. Moreover, since $A_q$ and
$Y_q$ have the same $\H$-charges, the dependence on $\Phi$, apart from
different (real) numerical coefficients, must be the same. This is the
key point in solving the SUSY CP problem. Consider indeed a simple one
flavor case. If the combination of horizontal symmetry and holomorphy
allows only one combination of flavon fields $\Phi_1$ to contribute to
$A$ and $Y$, then we have
\eqn\onedim{Y= y\Phi_1,\quad\quad A=m \Phi_1\ \ \
\Longrightarrow\ \ \ \phi_A=\arg\left({m\over y}\right)=0,}
even when $\Phi_1$ is complex ($y$ and $m$ are real parameters according
to the assumption of a CP conserving Lagrangian). This mechanism would
fail if two combinations of fields, $\Phi_1$ and $\Phi_2$, contributed:
\eqn\onedimcp{Y=y_1\Phi_1+y_2\Phi_2,\quad\quad A=m_1\Phi_1+m_2\Phi_2\ \ \
\Longrightarrow\ \ \  \phi_{A}={\rm arg}\left({m_1\Phi_1+m_2\Phi_2\over
y_1\Phi_1+y_2\Phi_2}\right )\neq 0}
(for $\Im(\Phi_1\Phi_2^*)\neq0$ and $y_1/y_2\neq m_1/m_2$, which is
the general case). This analysis is easily generalized to the
relevant case where $A$ and $Y$ are $3\times3$ matrices which have
complex entries in order to generate a non-zero CKM phase. The result is
that SUSY CP phases are suppressed when each eigenvalue in $Y$ is
determined to high accuracy by just one combination of operators
(also including contributions from off-diagonal terms).

In the model of ref. \RaNi, as a result of the flavor symmetry and
holomorphy, the form of the quark Yukawa matrices is approximately
triangular. The suppression of multiple contributions to the eigenvalues
is mainly due to that.  The effective phases are $\O(\lambda^6)$,
leading to $d_N\sim 10^{-28}e\ {\rm cm}$.

In models of alignment \refs{\QSA,\LNSb}, in order to obtain the Cabibbo
angle as experimentally measured, it is necessary that the supersymmetric
mixing angle between $\tilde u_L-\tilde c_L$ is  $\O(\theta_C)$. This
leads to $D-\bar D$ mixing close to the experimental bound. Furthermore,
with an arbitrary new CP violating phase in the mixing matrix,
interesting CP violating effects are likely to arise, {\it i.e.}
a different time dependence between the rates of $D^0\ra K^+\pi^-$ and
$\bar D^0\ra K^-\pi^+$
\ref\BSN{G. Blaylock, A. Seiden and Y. Nir,
 Phys. Lett. B355 (1995) 555.}.
It is, however, possible that the mechanism that solves the SUSY CP
problems also constrains the new CP effects in $D-\bar D$ mixing to be
negligible (as it is indeed the case in the model of ref. \RaNi).

For the neutral $B$ system the relevant supersymmetric mixing angles
are suppressed  by $\O(V_{ub})$. The supersymmetric
contribution to $B-\bar B$ mixing can be comparable to the SM
contribution for squark masses around $300\ \gev$  \LNSb. The crucial
difference with respect to exact universality does not lie, however, in
the magnitude of the contributions: these may be too small to be clearly
signaled in $\Delta m_B$ because of the hadronic uncertainties
(most noticeably in $f_B$). It lies instead in the fact that the phase of
the supersymmetric contribution  is now generically different from that
of the standard $W$-boson box diagrams. Therefore, in models where
Abelian flavor symmetries tame the supersymmetric FCNC, large deviations
from the SM in CP asymmetries in neutral $B$ decays are possible.

\subsec{The Non-Universal Case: Approximate Non-Abelian Horizontal
Symmetries}

\nref\PoTo{A. Pomarol and D. Tommasini, Nucl. Phys. B466 (1996) 3.}%
\nref\DKL{M. Dine, A. Kagan and R.G. Leigh, Phys. Rev. D48 (1993) 4269.}%

In this section, we discuss a mechanism which we call {\it approximate
universality}, and which is mainly devised to solve (though, in most
models, it only relaxes) the $\epsK$ problem. This mechanism is typically
associated with models of non-Abelian horizontal symmetries, $\H$, where
quarks of the two light families fit into an irreducible doublet. In the
flavor basis one expects the splitting among the squarks of these
families to be  $\O(\Phi\Phi^*/\Lambda_F^2)$, where $\Phi$ breaks $\H$
and $\Lambda_F$ is the flavor scale. The ratio $\lambda^2=\Phi\Phi^*/
\Lambda_F^2$ is expected to be of the order of some products of CKM
mixing angles or light to heavy quark mass ratios \DKL\ (typically
$\lambda\sim\thc\sim0.2$), leading to a suppression of $\epsK$.

Let us discuss in more detail this mechanism by focusing on the $(1,2)$
family sector
\ref\RaSU{Y. Nir and R. Rattazzi, CERN-TH/96-350 (1996).}.
In the flavor basis, the $2\times2$ Yukawa matrices have the form
\eqn\twobytwo{Y^d=Y_s\pmatrix{y_{11}^d&y^d_{12}\cr y^d_{21}&1\cr},\ \ \
Y^u=Y_c\pmatrix{y_{11}^u&y^u_{12}\cr y^u_{21}&1\cr}.}
One motivation for models of non-Abelian horizontal symmetries
is that they can give $y^d_{12}=y^d_{21}\simeq\thc$ and
$y^d_{11}\ll y^d_{12}y^d_{21}$. Then,
since the CKM mixing is mostly generated in the down sector, the well
known successful prediction $|V_{us}|\simeq\sqrt{m_d/m_s}$ is generated.

By phase rotations it is always possible to choose $y_{12}^d,y_{21}^d$
real, while $y_{11}^d$ and $y^u_{12}$ are, in general, complex.
We define then two CP violating phases:
\eqn\defphaNA{\psi_d={\Im\ y_{11}^d\over y_{12}^dy_{21}^d},\ \ \
\psi_u={\Im \ y_{12}^u\over y_{12}^d}.}
The phase $\psi_d$ is the phase of the $d$-quark Yukawa coupling, while
$\psi_u=\arg(V_{us})$. In this convention the (charged current)
electroweak hamiltonian for kaon decays is complex with a phase $\psi_u$.
In the usual convention, where the electroweak hamiltonian is real,
$\psi_u$ will appear in the $\Delta S=2$ amplitude.
We further define the mass-squared splitting between the
diagonal entries of the squark mass matrices in the flavor basis,
\eqn\split{\delta_{Q,D}=\delta m_{Q,D}^2/\tilde m^2.}
In most models, the dominant supersymmetric contribution to $\epsK$ is
proportional to
\eqn\flavorbasis{\Im\left\{{(\delta m_Q^2)_{12}\over m_Q^2}
{(\delta m_D^2)_{12}\over m_D^2}\right\}\simeq y^d_{12}y^d_{21}
\delta_Q\delta_D(\psi_d+\psi_u).}
Notice that, generically, $\psi_d$ contributes to $d_N$, so that we
expect $\psi_d\lsim10^{-2}$ independent of squark splittings.
On the other hand, in the simplest cases, $\lambda\sim0.2$ and
$\delta_Q\delta_D\sim\lambda^4\sim 10^{-3}$, so that eq. \epsKScon\ gives
a somewhat stronger bound, $\psi_d\lsim 10^{-3}$. Nonetheless, a
mechanism to suppress $d_N$ will also suppress this contribution to
$\epsK$, possibly by a sufficient amount. In particular, there are
interesting models with $y_{11}^{u,d}=0$, so that $\psi_d=0$.
With regard to $\psi_u$, the situation is more problematic. Typically,
$y^u_{12}\sim\sqrt{m_u/m_c}\sim\lambda^2$ with an arbitrary phase,
so that $\psi_u=\O(\lambda)$. This result is actually unavoidable in the
interesting class of models where the (13) and (11) entries of the Yukawa
matrices vanish in the flavor basis, leading to the predictions
$|V_{td}/V_{ts}|\simeq \sqrt{m_d/m_s}$ and $|V_{ub}/V_{cb}|\simeq
\sqrt{m_u/m_c}$. To obtain a large CKM phase, a large
$\psi_u$ is necessary. Consequently, the $\epsK$-bound becomes
$\delta_Q\delta_D\lsim 10^{-5}\sim\lambda^7$. Many simple models
\nref\PoSe{P. Pouliot and N. Seiberg, Phys. Lett. B318 (1993) 169.}%
({\it e.g.} \refs{\PoTo,\PoSe}) have $\delta_Q\delta_D\sim\lambda^4$
which relaxes but does not completely solve the $\epsK$ problem.
A similar situation holds in other models which do satisfy
$\delta_Q\delta_D\lsim 10^{-5}$, but which generate somewhat
bigger effects via  the mixing with the third family
\ref\BDH{R. Barbieri, G. Dvali and L.J. Hall,
 Phys. Lett. B377 (1996) 76.}\
or via $[(\delta m_Q^2)_{12}]^2$
\ref\HaMu{L.J. Hall and H. Murayama, Phys. Rev. Lett. 75 (1995) 3985.}.
There exist however specific models
\nref\CHM{C. Carone, L.J. Hall and H. Murayama,
 Phys. Rev. D54 (1996) 2328.}%
\nref\Raby{R. Barbieri, L.J. Hall, S. Raby and A. Romanino,
hep-ph/9610449.}%
\refs{\CHM-\Raby} where the non-Abelian symmetry does solve the $\epsK$
problem completely.

We emphasize, nonetheless, that relaxing (without completely solving) the
problem is still useful. Essentially all these models would be acceptable
if  the level of degeneracy were a factor of $10$ stronger than
the naive expectation from $\H$ selection rules. It has been suggested
that a stronger degeneracy may be dynamically induced by RG evolution
\LaRo. Gluino dominance in the squark mass evolution is a possible
mechanism, since the contribution of gluino masses through RG evolution
is universal. For $m_Q^2\sim m_{\tilde g}^2$ at the Planck scale,
the gluino contribution to the low energy squark mass-squared dominates
the overall original one by a factor of $6\div 7$. This additional
degeneracy is just about what is needed. Other possibilities to
completely solve the $\epsK$ problem in these models, without increasing
the level of degeneracy, are an approximate CP symmetry (see section
2.4 above) or heavy squarks (see section 2.7 below).

We would like to add a few  comments on the SUSY CP problem in models
with approximate universality. The same mechanism of screening discussed
in the previous section may work for models of non-Abelian symmetries.
However, the more constrained form of the Yukawa matrices, and in
particular the non-zero (12) and (21) entries, generally leads to a
weaker suppression of SUSY phases \RaSU. Consider for instance the class
of models, which includes ref. \RaNi, where the CKM phase originates
from some off-diagonal entry in the mass matrix which receives
contributions from more than one combination of flavon fields
with a relative non-trivial phase.
In general, it can be shown that non-universality of $A$-terms and
the requirement of $\O(1)$ CKM phase imply $\phi_A\gsim
\lambda^6\sim J$. The minimal result can be reached only if $Y_{21}$
is highly suppressed (or vanishing), which can be achieved with
Abelian flavor symmetries. In models of non-Abelian symmetries, where
the two light families are in irreducible doublets, one does not expect
this suppression of $Y_{21}$ to hold, so that there are more
contributions to the eigenvalues of the light quarks. For example, when
$|Y_{12}|\sim|Y_{21}|\sim \sqrt {|Y_{11}Y_{22}|}$, the effective CP
phases for light quarks are expected to be $\gsim\lambda^4$. If, in
addition, $|Y_{13}|\sim|Y_{31}|$, it is difficult to avoid an effective
phase $\sim \lambda^2$, {\it i.e.} barely compatible with present bounds
(see, for example,
\ref\BHS{R. Barbieri, L.J. Hall and A. Strumia,
 Nucl Phys. B449 (1995) 437.}\
for a discussion of GUT scenarios).

The situation in supersymmetric models without flavor universality is
then very interesting. On one side, we should not be surprised that
$d_N$ lies below the experimental bound. While the models contain
new CP violating phases, they also provide a mechanism, directly
connected to holomorphy (see eq. \onedim), to screen the new CP phases.
On the other side,
since $\delta_{\rm KM}(\sim 1)$ feeds into SUSY phases in a much less
suppressed way than in the MSSM (see eq. \gim), there should be a
non-negligible amount of CP violation in $A$-terms. In a large class of
models this leads to $d_N\gsim 10^{-28}e\ {\rm cm}$, with the minimum
corresponding to $\phi_A^d\sim \lambda^6$ and obtained for specific
textures. It is encouraging that this minimal $d_N$ seems to be within
the reach of the next generation of experiments (see, for example,
\ref\GoLa{R. Golub and K. Lamoreaux, Phys. Rep. 237 (1994) 1.}).
While this is about three orders of magnitude below the present
experimental bound, it is still a few orders of magnitude above the SM
value and, for that matter, the value in the MSSM with exact
universality.

In models with approximate universality, the expected size of
$D-\bar D$ mixing is at least 2-3 orders of magnitude below the present
bound. For processes involving the third family, such as $B-\bar B$
mixing, non-Abelian models with the third family in a singlet of $\H$
have signatures similar to those of Abelian models. Therefore, similarly
to models of alignment, large deviations from the
SM in CP asymmetries in neutral $B$ decays are possible.

\subsec{Heavy Squarks}

The Supersymmetric CP problem is solved and the $\epsK$ problem is
relaxed (but not eliminated)
if the masses of the first and second generations squarks $m_i$
are larger than the other soft masses, $m_i^2\sim 100\, \tilde m^2$
\refs{\PoTo,\DKL}. This does not necessarily lead to naturalness
problems, since these two generations are almost decoupled from the Higgs
sector. Explicit models are presented in
\nref\CKNS{A.G. Cohen, D.B. Kaplan and A.E. Nelson,
 Phys. Lett. B388 (1996) 588.}%
\nref\DvaPo{G. Dvali and A. Pomarol, CERN-TH-96-192, hep-ph/9607383.}%
\refs{\PoTo,\CKNS-\DvaPo}.

Notice though that, with the possible exception of $m_{\tilde b_R}^2$,
third family squark masses cannot naturally be much above $m_Z^2$. Then
for non-zero CP phases in this sector (or for $\phi_B\neq0$) one can
still generate a sizeable EDM of the neutron via the two-loop induced
three-gluon operator
\ref\Weintg{S. Weinberg, Phys. Rev. Lett. 63 (1989) 2333.}.
Indeed, for a light right-handed sbottom, the contribution to $d_N$ is
about \FPT
\eqn\lightsb{d_N\sim\left({100{\rm GeV}\over m_{\tilde b}}\right)^2
\phi_{A_b}\times 10^{-24}\ e\ {\rm cm} .}
For top squarks, naturalness constrains both stops to be light, but their
contribution is about an order of magnitude smaller because of the
different QCD dressing
\nref\BLY{E. Braaten, C.S. Li and T.C. Yuan,
 Phys. Rev. Lett. 64 (1990) 1709.}%
\nref\BGTW{G. Boyd, A.K. Gupta, S.P. Trivedi anf M.B. Wise,
 Phys. Lett.  B241 (1990) 584.}%
\nref\DDD{J. Dai, H. Dykstra, R.G. Leigh, S. Paban and D. Dicus,
 Phys. Lett. B237 (1990) 216, (E) B242 (1990) 547.}%
\nref\DiFi{M. Dine and W. Fischler, Phys. Lett. B242 (1990) 239.}%
\refs{\BLY-\DiFi}. We conclude that, if phases are generically of order
one, the main contribution to $d_N$ comes from the
third family and it is roughly at the present experimental bound
when $m_{\tilde t_{L,R}}\sim 100\ \gev$.

The upper bound from naturalness on the first two generations is
$m_{Q,D}\lsim20\ TeV$ for low $\Lambda_S$, and even
stronger, $2-5\ \tev$, for $\Lambda_S\sim M_{\rm Pl}$
\ref\GiuDim{S. Dimopoulos and G.F. Giudice,
 Phys. Lett. B357 (1995) 573.}.
When these bounds are taken into account, eq. \epsKScon\ is not
satisfied, in general. Combining this scenario with alignment,
$(\delta m_{Q,D}^2)_{12}\sim\sin\theta_C\ \delta m_{Q,D}^2$, would solve
the $\Delta m_K$ problem, but the contribution to $\epsK$ would still be
too large, unless
\eqn\HeaAli{{\delta m_Q^2\over m_Q^2}{\delta m_D^2\over m_D^2}\sin\phi
\lsim 10^{-2}.}
However, this scenario becomes viable when further combined
with the approximate
universality of the models with non-Abelian horizontal symmetries.
All those models which were problematic with light squarks do satisfy the
milder eq. \HeaAli. Notice, though, that the universal contribution to
squark masses from gluino terms in the RG evolution cannot play a
significant role here. This is because gluino masses affect the stop
mass and are thus constrained by naturalness to be around the weak scale.

Models with the first two squark generations heavy
have their own signatures of CP violation in neutral meson mixing
\ref\CKLN{A.G. Cohen, D.B. Kaplan, F. Lepeintre and A.E. Nelson,
 Phys. Rev. Lett. 78 (1997) 2300.}.
The mixing angles relevant to $D-\bar D$ mixing are similar, in general,
to those of models of alignment (if alignment is
invoked to explain $\Delta m_K$ with $m^2_{Q,D}\lsim20\ TeV$).
However, as $\tilde u$ and $\tilde c$ squarks are heavy, the
contribution to $D-\bar D$ mixing is only about one to two orders
of magnitude below the experimental bound. This may lead to the
interesting situation that $D-\bar D$ mixing will first be observed
through its CP violating part
\ref\WolfD{L. Wolfenstein, Phys. Rev. Lett. 75 (1995) 2460.}.
In the neutral $B$ system, $\O(1)$ shifts from the Standard Model
predictions of CP asymmetries in the decays to final CP eigenstates
are possible. This can occur even when the squarks masses of the third
family are $\sim1\ \tev$ \CKNS, since now mixing angles can naturally
be larger than in the case of horizontal symmetries
(alignment or approximate universality).

\newsec{Extensions of the Fermion Sector}

The fermion sector of the Standard Model consists of three
generations, with ($i=1,2,3$)
\eqn\gens{Q_i(3,2)_{+1/6},\ \ \bar u_i(\bar 3,1)_{-2/3},\ \
\bar d_i(\bar 3,1)_{+1/3},\ \
L_i(1,2)_{-1/2},\ \ \bar\ell_i(1,1)_{+1},}
representations of the $SU(3)\times SU(2)\times U(1)$ gauge group.
It can be extended by either a fourth, sequential generation or
by non-sequential fermions, namely `exotic' representations,
different from those of \gens.\foot{The four generation model became
rather unlikely in view of the experimental fact that there are
only three massless (or light) neutrinos from $L_i$
representations. However, if neutrinos acquire their masses from a
see-saw mechanism, and if the scale of right-handed neutrino masses
is close to the electroweak one, then it is quite possible that
a fourth generation neutrino is heavy enough to evade experimental
and cosmological constraints
\ref\HaNi{H. Harari and Y. Nir, Nucl. Phys. B292 (1987) 251.}.}
Most of our discussion in this chapter is focused
on non-sequential fermions and their implications on CP asymmetries
in neutral $B$ decays. (For the general formalism of CP asymmetries
in neutral $B$ decays see {\it e.g.}
\nref\NirRev{Y. Nir, Lectures presented in the 20th SLAC Summer
Institute, SLAC-PUB-5874 (1992).}%
\nref\NQRev{Y. Nir and H.R. Quinn,
 Ann. Rev. Nucl. Part. Sci. 42 (1992) 211.}%
\nref\DunRev{I. Dunietz, Ann. Phys. 184 (1988) 350.}%
\refs{\NirRev-\DunRev}. For model-independent analyses of New Physics
effects see
\nref\NiSigen{Y. Nir and D. Silverman, Nucl. Phys. B345 (1990) 301.}%
\nref\NiQugen{Y. Nir and H.R. Quinn, Phys. Rev. D42 (1990) 1473.}%
\nref\DLNgen{C.O. Dib, D. London and Y. Nir,
 Int. J. Mod. Phys. A6 (1991) 1253.}%
\nref\DDO{N.G. Deshpande, B. Dutta and S. Oh,
 Phys. Rev. Lett. 77 (1996) 4499.}%
\nref\SiWogen{J.P. Silva and L. Wolfenstein, hep-ph/9610208.}%
\nref\GrWogen{Y. Grossman and M.P. Worah, Phys. Lett. B395 (1997) 241.}%
\refs{\NiSigen-\GrWogen}.) The last two sections discuss other processes:
$K_L\to\pi\nu\bar\nu$ (with non-sequential quarks)
and $\Delta\Gamma(B_s)$ (with  a sequential fourth generation).

\subsec{The Theoretical Framework}

We consider a model with extra quarks in vector-like
representations of the standard Model gauge group,
\eqn\extra{d_4(3,1)_{-1/3}\ +\ \bar d_4(\bar 3,1)_{+1/3},}
Such (three pairs of) quark representations appear, for example, in $E_6$
GUTs. The most interesting effects in this model
concern CP asymmetries in neutral $B$ decays into final CP eigenstates
\nref\NiSiZ{Y. Nir and D. Silverman, Phys. Rev. D42 (1990) 1477.}%
\nref\SilvZ{D. Silverman, Phys. Rev. D45 (1992) 1800.}%
\nref\BMPRZ{G.C. Branco, T. Morozumi, P.A. Parada and M.N. Rebelo,
 Phys. Rev. D48 (1993) 1167.}%
\nref\SilvZ{W-S. Choong and D. Silverman, Phys. Rev. D49 (1994) 2322.}%
\nref\BBPZ{V. Barger, M.S. Berger and R.J.N. Phillips,
 Phys. Rev. D52 (1995) 1663.}%
\nref\SilZ{D. Silverman, Int. J. Mod. Phys. A11 (1996) 2253.}%
\nref\GrLo{M. Gronau and D. London, Phys. Rev. D55 (1997) 2845.}%
\refs{\NiSiZ-\GrLo}. We describe these effects in detail as they
illustrate the type of new ingredients that are likely to affect
CP asymmetries in neutral $B$ decays and the way in which the SM
predictions might be modified.\foot{If there exist light up quarks in
exotic representations, they may introduce similar, interesting
effects in neutral $D$ decays \BSN.}

The most important feature of this model for our purposes is that
it allows CP violating $Z$-mediated Flavor Changing Neutral Currents
(FCNC). To understand how these FCNC arise, it is convenient to work
in a basis where the up sector interaction eigenstates are identified
with the mass eigenstates. The down sector interaction eigenstates
are then related to the mass eigenstates by a $4\times4$ unitary
matrix $K$. Charged current interactions are described by
\eqn\Zcci{\eqalign{
\L^W_{\rm int}\ =&\ {g\over\sqrt2}(W_\mu^-J^{\mu+}+W_\mu^+J^{\mu-}),\cr
J^{\mu-}\ =&\ V_{ij}\bar u_{iL}\gamma^\mu d_{jL}.\cr}}
The charged current mixing matrix $V$ is a $3\times4$ sub-matrix of $K$:
\eqn\ZVinK{V_{ij}=K_{ij}\ \ {\rm for}\ \ i=1,2,3;\ j=1,2,3,4.}
The $V$ matrix is parameterized by six real angles and {\it three}
phases, instead of three angles and one phase in the original CKM matrix.
All three phases may affect CP asymmetries in $B^0$
decays. Neutral current interactions are described by
\eqn\Zcci{\eqalign{
\L_{\rm int}^Z\ =&\ {g\over\cos\theta_W}Z_\mu
(J^{\mu3}-\sin^2\theta_WJ^{\mu}_{\rm EM}),\cr
J^{\mu3}\ =&\ -{1\over2}U_{pq}\bar d_{pL}\gamma^\mu d_{qL}
+{1\over2}\delta_{ij}\bar u_{iL}\gamma^\mu u_{jL}.\cr}}
The neutral current mixing matrix for the down sector is $U=V^\dagger V$.
As $V$ is not unitary, $U\neq{\bf 1}$. In particular, its non-diagonal
elements do not vanish:
\eqn\ZUpq{U_{pq}=-K_{4p}^*K_{4q}\ \ {\rm for}\ \ p\neq q.}
The three elements which are most relevant to our study are
\eqn\Zrele{\eqalign{
U_{ds}\ =&\ V_{ud}^*V_{us}+V_{cd}^*V_{cs}+V_{td}^*V_{ts},\cr
U_{db}\ =&\ V_{ud}^*V_{ub}+V_{cd}^*V_{cb}+V_{td}^*V_{tb},\cr
U_{sb}\ =&\ V_{us}^*V_{ub}+V_{cs}^*V_{cb}+V_{ts}^*V_{tb}.\cr}}
The fact that, in contrast to the Standard Model, the various $U_{pq}$
do not necessarily vanish, allows FCNC at tree level. This may
substantially modify the predictions for CP asymmetries.

\subsec{Implications of $Z$-Mediated FCNC}

The flavor changing couplings of the $Z$ contribute to various FCNC
processes. Relevant constraints arise from semileptonic FCNC $B$ decays:
\eqn\ZBll{{\Gamma(B\ra\ell \bar \ell X)_Z\over\Gamma(B\ra e\nu X)}=
\left[(T_3-Q\sin^2\theta_W)^2+(Q\sin^2\theta_W)^2\right]
{|U_{db}|^2+|U_{sb}|^2\over|V_{ub}|^2+F_{\rm ps}|V_{cb}|^2},}
where $F_{\rm ps}\sim0.5$ is a phase space factor,
$\ell$ is any of $\nu_i$ ($i=1,2,3$), $e^-$ or $\mu^-$,
$T_3=+1/2[-1/2]$ and $Q=0[-1]$ for $\nu_i[\ell^-]$. The experimental
upper bounds
\nref\UAone{C. Albajar {\it et al.}, UA1 Collaboration,
 Phys. Lett. B262 (1991) 163.}%
\nref\GLN{Y. Grossman, Z. Ligeti and E. Nardi,
Nucl. Phys. B465 (1996) 369,
(E) B480 (1996) 753.}%
\nref\bsnnAleph{ALEPH Collaboration, Report no. PA10-019, presented at
the 28th International Conference on High Energy Physics,
Warsaw, Poland (1996).}%
\refs{\UAone-\bsnnAleph}\ and, in particular, a preliminary D0 result
\ref\DzeroBxll{
D0 collaboration, Report no. FERMILAB-CONF-96/253-E,
presented at the 28th International Conference on High Energy Physics,
Warsaw, Poland (1996).},
BR$(B\to X\mu^+\mu^-)\leq3.6\times10^{-5}$, imply then
\eqn\ZUqb{\left|{U_{db}\over V_{cb}}\right|\leq0.04,\ \ \
\left|{U_{sb}\over V_{cb}}\right|\leq0.04.}
Additional constraints come from neutral $B$ mixing:
\eqn\ZdelmB{(x_d)_Z={\sqrt2G_FB_Bf_B^2M_B\eta_B\tau_b\over6}
|U_{db}|^2.}
The resulting bound is sensitive to the range taken for the
poorly known parameter $f_B$. It is of order $|U_{db}|\lsim10^{-3}$
which is comparable to \ZUqb. As for $x_s$, only lower bounds exist
and consequently there is no analog bound on $|U_{sb}|$.

If the $U_{qb}$ elements
are not much smaller than the bounds \ZUqb, they will affect several
aspects of physics related to CP asymmetries in $B$ decays.

(i) Neutral $B$ mixing:

The experimentally measured value of $x_d$ (and the lower bound on
$x_s$) can be explained by Standard Model processes, namely box
diagrams with intermediate top quarks. Still, the uncertainties in
the theoretical calculations, such as the values of $f_B$ and $V_{td}$
(and the absence of an upper bound on $x_s$) allow a situation where
SM processes do not give the dominant contributions to either or both
of $x_d$ and $x_s$. For example, for $m_t\approx180\ GeV$,
\eqn\lowSM{(x_d)_{\rm box}=0.17
\left[{\sqrt{B_B}f_B\over0.14\ GeV}\right]^2
\left[{\tau_b|V_{cb}|^2\over3.5\times10^9\ GeV^{-1}}\right]
\left[{|V_{td}/V_{cb}|\over0.12}\right]^2,}
namely, the Standard Model box diagrams could account for as little
as 25\% of the experimental value of $x_d$, and even less if the
unitarity of the CKM matrix does not hold, in which case the lower bound
$|V_{td}/V_{cb}|\geq0.12$ can be violated.

The ratio between the $Z$-mediated tree diagram and the Standard
Model box diagram is given by  ($q=d,s$)
\eqn\ZtoSM{{(x_q)_{\rm tree}\over(x_q)_{\rm box}}={\sqrt2\pi^2
\over G_F m_W^2 y_t f_2(y_t)}\left|{U_{qb}\over V_{tq}V_{tb}^*}\right|^2
\approx80\left|{U_{qb}\over V_{tq}V_{tb}^*}\right|^2.}
{}From \ZUqb\ and \ZtoSM\ we learn that the $Z$-mediated
tree diagram could give the dominant contribution to $x_d$
but at most $\O(0.1)$ of $x_s$.

(ii) Unitarity of the $3\times3$ CKM matrix.

Within the SM, unitarity of the three generation CKM matrix gives:
\eqn\three{\eqalign{
\U_{ds}\equiv\ &\ V_{ud}^*V_{us}+V_{cd}^*V_{cs}+V_{td}^*V_{ts}=0,\cr
\U_{db}\equiv\ &\ V_{ud}^*V_{ub}+V_{cd}^*V_{cb}+V_{td}^*V_{tb}=0,\cr
\U_{sb}\equiv\ &\ V_{us}^*V_{ub}+V_{cs}^*V_{cb}+V_{ts}^*V_{tb}=0.\cr}}
Eq. \Zrele, however, implies that now \three\ is replaced by
\eqn\reprel{\U_{ds}=U_{ds},\ \ \ \U_{db}=U_{db},\ \ \ \U_{sb}=U_{sb}.}
A measure of the violation of \three\ is given by
\eqn\viothr{
\left|{U_{ds}\over V_{ud}V_{us}^*}\right|\,\leq \, 5 \times 10^{-4},\ \ \
\left|{U_{db}\over V_{td}V_{tb}^*}\right|\,\leq \,0.3,\ \ \
\left|{U_{sb}\over V_{ts}V_{tb}^*}\right|\,\leq \,0.04.}
The bound on $|U_{db}/V_{td}V_{tb}^*|$ is even weaker if $|V_{td}|$ is
lower than the three generation unitarity bound. The bound on $|U_{ds}|$
follows from the experimental values of (or bounds on)
BR($K^+ \ra \pi^+ \nu \bar \nu$), $\epsilon_K$ and BR($K_L\ra\mu^+\mu^-$)
that we present later. The first of the SM relations in \three\ is
practically maintained, while the second can be violated by at most 5\%.
However, the $\U_{db}=0$ constraint may be violated by $\O(0.3)$ effects.
The Standard Model unitarity triangle should be replaced by a unitarity
{\it quadrangle}. A geometrical representation of the new relation
is given in fig. 1.

\epsfbox{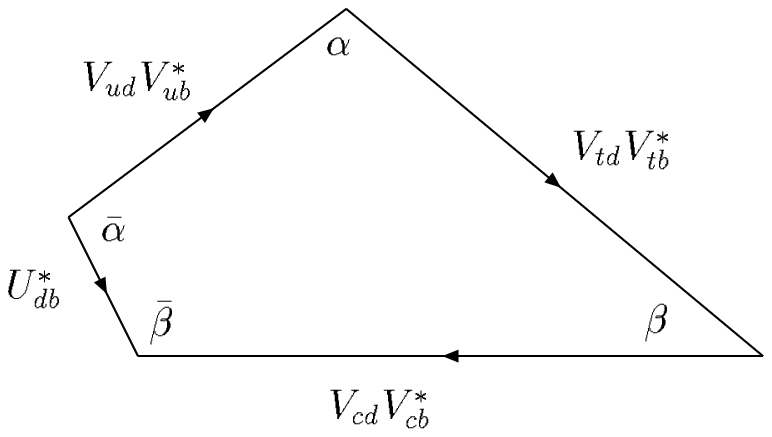}
\vbox{%
{\narrower\noindent%
\multiply\baselineskip by 3%
\divide\baselineskip by 4%
{\rm Figure 1. }{The  unitarity quadrangle
\medskip}}}

It should be stressed that, at present, only
the magnitudes of $U_{db}$ and $U_{sb}$ are constrained. Each of the
phases $\bar\alpha$ and $\bar\beta$,
\eqn\defbeal{\bar\alpha\equiv\arg\left({V_{ud}V_{ub}^*\over U_{db}^*}
\right),\ \ \ \
\bar\beta\equiv\arg\left({U_{db}^*\over V_{cd}V_{cb}^*}\right),}
could be anywhere in the range $[0,2\pi]$.

(iii) $Z$-mediated $B$ decays.

Our main interest in this chapter is in hadronic $B^0$ decays to
CP eigenstates, where the quark sub-process is $\bar b\ra\bar u_i
u_i\bar d_j$, with $u_i=u,c$ and $d_j=d,s$. These decays get
new contributions from $Z$-mediated tree diagrams, in addition
to the standard $W$-mediated ones. The ratio between the amplitudes is
\eqn\ratamp{{A_Z\over A_W}=\left[{1\over2}-{2\over3}\sin^2\theta_W
\right]\left|{U_{jb}^*\over V_{ij}V_{ib}^*}\right|.}
We find that the $Z$ contributions
can be safely neglected in $\bar b\ra \bar cc\bar s$ ($\lsim0.013$) and
$\bar b\ra\bar cc\bar d$ ($\lsim0.06$).
On the other hand, it may be significant in
$\bar b\ra\bar uu\bar d$ ($\lsim0.25$), and in processes with
no SM tree contributions, e.g. $\bar b\ra\bar ss\bar s$, that may have
comparable contributions from penguin and $Z$-mediated tree diagrams.

(iv) New contributions to $\Gamma_{12}(B_q)$

The difference in width comes from modes that are common to $B_q$ and
$\bar B_q$. As discussed above, there are new contributions to such modes
from $Z$-mediated FCNC. However, while the new contributions to $M_{12}$
are from tree level diagrams, {\it i.e.} $\O(g^2)$, those to $\Gamma_{12}$
are still coming form a box-diagram, {\it i.e.} $\O(g^4)$. Consequently,
no significant enhancement
of the SM value of $\Gamma_{12}$ is expected, and the relation
$\Gamma_{12}\ll M_{12}$ is maintained.\foot{The new contribution could
significantly modify the leptonic asymmetry in neutral $B$ decays
\ref\BPM{G.C. Branco, P.A. Parada and T. Morozumi,
 Phys. Lett. B306 (1993) 398.}, though the asymmetry remains small.}

\subsec{CP Asymmetries in $B$ Decays}

The fact that $M_{12}(B^0)$ could be dominated by the $Z$-mediated
FCNC together with the fact that this new amplitude depends
on new CP violating phases means that large deviations from the
Standard Model predictions for CP asymmetries are possible.
As $\Gamma_{12}\ll M_{12}$ is maintained, future measurements
of certain modes will still be subject to a clean theoretical
interpretation in terms of the extended electroweak sector parameters.

Let us assume that, indeed, $M_{12}$ is dominated by the new
physics.\foot{Generalization to the case that the new contribution is
comparable to (but not necessarily dominant over) the Standard Model
one is straightforward \BMPRZ.} Then
\eqn\Zmix{\left({p\over q}\right)_B\approx{U_{db}^*\over U_{db}}.}
We argued above that $b\ra c\bar cs$ is still dominated by the
$W$ mediated diagram. Furthermore, the first unitarity constraint in
\three\ is practically maintained. Then it is straightforward to evaluate
the CP asymmetry in $B\ra\psi K_S$. We find that it simply measures
an angle of the unitarity quadrangle of fig. 1:
\eqn\ZpsiKS{a_{CP}(B\ra\psi K_S)=-\sin2\bar\beta.}
The new contribution to $b\ra c\bar cd$ is  $\O(5\%)$,
which is the same order as the SM penguins and the expected
experimental sensitivity.
So we still have (taking into account CP-parities)
\eqn\Zsd{a_{CP}(B\ra\psi K_S)\approx-a_{CP}(B\ra DD).}
Care has to be taken regarding $b\ra u\bar ud$ decays. Here, direct
CP violation may be large \GrLo\ and prevent a clean theoretical
interpretation of the asymmetry. Only if the asymmetry is large, so
that the shift from the $Z$-mediated contribution to the decay is small,
we get
\eqn\Zpipi{a_{CP}(B\ra\pi\pi)=-\sin2\bar\alpha.}
The important point about the modification of the SM
predictions is then not that the angles $\alpha,\beta$ and $\gamma$
may have very different values from those predicted by the SM, but
rather that the CP asymmetries do not measure these angles anymore.
As there are no experimental constraints on $\bar\alpha$ and $\bar\beta$,
the full range $[-1,+1]$ is possible for each of the asymmetries.
This model demonstrates that there exist extensions of the SM
where dramatic deviations from its predictions for CP asymmetries
in $B$ decays are not unlikely.

Another interesting point concerns $B_s$ decays. As $B_s-\bar B_s$
mixing as well as the $b\ra c\bar cs$ decay are dominated by the SM
diagrams, we have, as in the SM,
\eqn\Bspsiphi{a_{CP}(B_s\ra\psi\phi)\approx0.}
As shown in ref. \NiSigen,
this is a sufficient condition for the angles extracted from
$B\ra\psi K_S$, $B\ra\pi\pi$ and $B_s\ra\rho K_S$ to sum up to $\pi$
(up to possible effects of direct CP violation). This happens in
spite of the fact that the first two asymmetries do not correspond
to $\beta$ and $\alpha$ of the unitarity triangle.

\subsec{The $K_L \to \pi \nu \bar \nu$ Decay}

$K_L \to \pi \nu \bar \nu$ is dominated  by CP violating effects
\ref\Litt{L.S. Littenberg, Phys. Rev. D39 (1989) 3322.}.\foot{%
Significant CP conserving contributions can arise if lepton
flavor is violated. Otherwise, CP conserving contributions
are highly suppressed
\ref\Yuval{Y. Grossman and Y. Nir, SLAC-PUB-7380, hep-ph/9701313.}.}
In the SM, the decay amplitude is dominated by top penguin and box
diagrams and can be calculated with very little theoretical uncertainties
\ref\Buraskl{A.J. Buras, Phys. Lett. B333 (1994) 476.}.
It then provides a clean measurement of the CP violating measure $J$
and, together with $K^+ \to \pi^+ \nu \bar \nu$, of the angle $\beta$
\ref\BuBu{G. Buchalla and A.J. Buras, Phys. Rev. D54 (1996) 6782.}.
Studies of New Physics effects on this decay can be found in
\nref\BeGeTu{G. Belanger, C.Q. Geng, and P. Turcotte, Phys. Rev.
D46 (1992) 2950.}%
\nref\CDS{C.E. Carlson, G.D. Dorata and M. Sher,
Phys. Rev. D54 (1996) 4393.}%
\refs{\BeGeTu,\CDS}.
Below we present a model independent formalism to analyze this mode and
explain why it is a manifestation of CP violation in interference between
mixing and decay (which is the reason for its theoretical cleanliness).

In the neutral $K$ system, the deviation of $|q/p|$ from unity is
experimentally measured (by the CP asymmetry in $K_L\to\pi\ell\nu$) and
is $\O(\Re\ \epsK)$, that is negligibly small for the purposes of this
section. For $|q/p|=1$, the time dependence of an untagged sample of $K$
mesons to decay into a CP eigenstates $f(=\pi^0\nu\bar\nu)$ is given by
\ref\DunBs{I. Dunietz, Phys. Rev. D52 (1995) 3048.}
\eqn\untag{\Gamma\left[f\left(t\right)\right]  ={\Gamma\left(K\ra
f\right)\over 2}\bigg\{(1+|\lambda|^2)\left(e^{-\Gamma_S t} +
e^{-\Gamma_L t} \right)+ 2\Re\lambda \left(e^{-\Gamma_S t}
-e^{-\Gamma_L t}\right)\bigg\},}
where we define
\eqn\deflam{\lambda \equiv {q \over p}{\bar A \over A}.}
The amplitude $A(\bar A$) is the $K^0(\bar K^0)\to f$ transition
amplitude. The deviation of the ratio $\bar A/A$ from unity is also
negligibly small. Since there is only one neutral hadron at the final
state, there is no final state interaction phase. An absorptive
phase can come from diagrams with two real intermediate pions, like those
arising at higher order in chiral perturbation theory, and is negligibly
small
\ref\GeHsLi{C.Q. Geng, I.J. Hsu and Y.C. Lin,
Phys. Lett. B355 (1995) 569.}.
Therefore, it is safe to assume that $|\lambda|=1$ to an excellent
approximation, and the leading CP violating effect is
then $\Im\lambda\neq0$, namely interference between mixing and decay.

Defining $\theta$ to be the relative phase between the $K-\bar K$
mixing amplitude and the $s\to d\nu\bar\nu$ decay amplitude, namely
$\lambda=e^{2i\theta}$, we get from \untag:
\eqn\genrat{
{\Gamma(K_L\to\pi^0\nu\bar\nu)\over\Gamma(K_S \to \pi^0 \nu \bar \nu)} =
{1 - \cos 2 \theta \over 1 + \cos 2 \theta} = \tan^2 \theta.}
This ratio measures $\theta$ without any information about the magnitude
of the decay amplitudes. In practice, however, it will be impossible to
measure $\Gamma(K_S\to\pi^0\nu\bar\nu)$. We can use the isospin symmetry
relation, $A(K^0\to\pi^0\nu\bar\nu)/A(K^+\to\pi^+\nu\bar\nu)=1/\sqrt{2}$,
and replace the denominator by the charged kaon mode:
\eqn\ratis{a_{CP} \equiv {\Gamma(K_L \to \pi^0 \nu \bar \nu) \over
\Gamma(K^+ \to \pi^+ \nu \bar \nu)} = r_{is}
{1 - \cos 2 \theta \over 2} = r_{is} \sin^2 \theta,}
where $r_{is} = 1.048$ is the isospin breaking factor
\ref\MaPa{W. Marciano and Z. Parsa, Phys. Rev. D53 (1996) 1.}.
The ratio \ratis\ may be experimentally measurable, as the relevant
branching ratios are $\O(10^{-10})$ in the SM and even larger in some of
its extensions. It will provide us with a very clean and
{\it model independent} measurement of the CP violating phase $\theta$.

New Physics can modify both the mixing and the decay amplitudes.
The contribution to the mixing can be of the same order as the SM one.
However, $\epsK=\O(10^{-3})$ implies that any new contribution
to the mixing amplitude carries the same phase as the SM one
(to $\O(10^{-3})$). On the other hand, the upper bound
\ref\Adler{S. Adler {\it et al.}, BNL 787 Collaboration,
 Phys. Rev. Lett. 76 (1996) 1421.}\
\eqn\expup{BR(K^+ \to \pi^+ \nu \bar \nu) < 2.4 \times 10^{-9},}
which is about 30 times larger than the SM prediction \BuBu, allows
New Physics to dominate the decay amplitude (with an arbitrary phase).
We conclude that the only potentially significant new contribution to
$a_{CP}$ can come from the decay amplitude. This is in contrast to the
clean CP asymmetries in the $B$ system where we
expect significant effects of New Physics only in the mixing amplitude.

$Z$-mediated FCNC provide an explicit example of New Physics that may
modify the SM prediction for $a_{CP}$ of eq. \ratis\ \Yuval.
Assuming that the $Z$-mediated tree diagram dominates
$K\ra\pi\nu\bar\nu$, we get \NiSiZ\
\eqn\Zkpnn{
{\Gamma(K^+\to\pi^+\nu\bar\nu)\over\Gamma(K^+ \to \pi^0 e^+ \nu )}=
r^+_{is}{1 \over 4} {|U_{ds}|^2\over|V_{us}|^2},\ \ \ \
{\Gamma(K_L\to\pi^0\nu\bar\nu)\over\Gamma(K^+ \to \pi^0 e^+ \nu )}=
r^0_{is}{1 \over 4} {|{\rm Im}U_{ds}|^2\over|V_{us}|^2}.}
Here $r^0_{is}=0.944$ and $r^+_{is}=0.901$ are the isospin breaking
corrections \MaPa\ (so that $r_{is}=r_{is}^0/r_{is}^+$). The ratio
\ratis\ measures, in this case, $\sin\theta=\Im U_{ds}/|U_{ds}|$.

Bounds on the relevant couplings come from $K_L\to\mu^+\mu^-$ (where we
take into account uncertainties from long distance contributions
\nref\Ko{P. Ko, Phys. Rev. D45 (1992) 174.}%
\nref\BucBu{G. Buchalla and A.J. Buras, Nucl. Phys. B412 (1994) 106.}%
\refs{\Ko-\BucBu}), from $K^+\to\pi^+\nu\bar\nu$ (see \ZBll\ and \expup),
and from the measurement of $\epsK$ \refs{\NiSiZ,\SilvZ}:
\eqn\sdreb{|\Re(U_{ds})| \lsim 2 \times 10^{-5},\ \ \
|U_{ds}| \leq 1.0 \times 10^{-4},\ \ \
|\Re(U_{ds}) \, \Im(U_{ds})| \lsim 1.3 \times 10^{-9}.}
We learn that large effects are possible. When $|\Re(U_{ds})|$ and
$|\Im(U_{ds})|$ are close to their upper bounds, the branching ratios
$BR(K^+ \to \pi^+ \nu \bar \nu)$ and $BR(K_L \to \pi^0 \nu \bar \nu)$
are both $O(10^{-9})$ and $a_{CP}$ of eq. \ratis\ is $\O(1)$.
Furthermore, as in this case the SM contribution is negligible,
the measurement of $BR(K^+ \to \pi^+ \nu \bar \nu)$ determines
$|U_{ds}|$, and with the additional measurement of
$BR(K_L \to \pi^0 \nu \bar \nu)$, $\arg(U_{ds})$ is determined as well.

To conclude, in models with lepton flavor conservation,
${\Gamma(K_L\to\pi^0\nu\bar\nu)\over\Gamma(K^+\to\pi^+\nu\bar\nu)}\neq0$
signifies CP violation. The value of this ratio provides a clean
measurement of a CP violating phase (possibly coming from New Physics).

\subsec{The $B_s$ Width Difference}

In the SM, a large width difference is expected in the $B_s$ system
\nref\hag{J.S. Hagalin, Nucl. Phys. B193 (1981) 123.}%
\nref\vuks{M.B. Voloshin, N.G. Uraltsev, V.A. Khoze and M.A. Shifman,
Sov. J. Nucl. Phys. 46, (1987) 112.}%
\nref\alek{R. Aleksan, A. Le Yaouanc, L. Oliver, O. P\`ene and J.-C.
Raynal, Phys. Lett. B316 (1993) 567.}%
\nref\BeBuDu{M. Beneke, G. Buchalla and I. Dunietz,
Phys. Rev. D54 (1996) 4419.}%
\refs{\hag-\BeBuDu}:
\eqn\BsSM{{\dg \over \Gamma} \approx 0.2.}
New physics can contribute significantly to the mass difference.
If this contribution is CP violating, it leads to a reduction of
the width difference
\ref\yuvalBs{Y. Grossman, Phys. Lett. B380 (1996) 99.}.
Below we explain this general result and give an explicit example:
the four generation model.

In general, the width difference is given by (for reviews, see e.g.
\refs{\NirRev-\DunRev})
\eqn\mastf{\dg ={4 \Re (M_{12} \Gamma^*_{12}) \over \dm} .}
The experimental lower bound $\dm/\,\Gamma > 8.8$
\ref\Bsexpbound{D. Buskulic {\it et al.}, ALEPH Collaboration,
 Phys. Lett. B356 (1995) 409.}\
implies $\dm \gg \dg$ and, consequently, $|M_{12}|\,\gg\,|\Gamma_{12}|$.
Thus, to a very good approximation, $\dm=2|M_{12}|$ and
\eqn\dgNP{\dg=2|\Gamma_{12}|\cos2\xi,\ \ \
2\xi\equiv\arg(-M_{12}\Gamma^*_{12}).}
Under the reasonable assumption that the New Physics does not
significantly affect the leading decay processes, $\Gamma_{12}$ is
dominated by $b\to c\bar cs$ transitions. Consequently, $2\xi$ is the
relative phase between the mixing amplitude and the $b \to c\bar cs$
decay  amplitude. In the SM,
\eqn\SMxi{\xi=\beta^\prime\equiv\arg\left(-{V_{cs}^*V_{cb}\over
V_{ts}V_{tb}^*}\right)\approx 0,}
and then $\cos2\xi=1$ to a very high accuracy. With new contributions to
the mixing amplitude, non-trivial phases may arise, leading to
$\cos2\xi<1$. This proves our statement: New CP violating contributions
to the mixing always reduce $\dg$ relative to the SM prediction.

As already mentioned, $Z$-mediated FCNC cannot contribute significantly
to the $B_s$ mass difference. However, the effects discussed above might
appear if there exists a fourth sequential generation.
The CKM matrix is extended to a unitary $4 \times4$ matrix, which can be
parameterized by 6 angles and 3 phases. There are new contributions to
$B_s$ mixing from box diagrams involving one or two $t^\prime$-quarks.
There are no experimental constraints that forbid the $t^\prime$
contribution to be comparable to or even dominate over the SM one. This
is the case if $|V_{t^\prime b}V_{t^\prime s}|$ is large.

The relevant effects are related to the modification of the
unitarity relation
\eqn\unirel{
\sum_i V_{ib}^* V_{is}^{} = 0,}
where $i$ runs over all up-type quarks. In the SM, the smallness of
$|V_{ub}^*V_{us}|$ leads to $\beta^\prime\approx0$ (see \SMxi).
With a fourth generation, the $i=c,t,t^\prime$ terms in \unirel\ can be
all of the same order. Then, both the SM phase and the new phase
from the $t^\prime$ contribution could be large:
\eqn\fourt{\arg\left({V_{cs}^*V_{cb}\over V_{ts}V_{tb}^*}\right) \neq 0,
\ \ \ \arg\left({V_{cs}^*V_{cb}\over V_{t^\prime s}V_{t^\prime b}^*}
\right) \neq 0.}
Consequently, $\cos 2\xi$ can assume any value and the $B_s$
width difference can be significantly smaller than in the SM.
Such a reduction is an indication of CP violation: the large SM
prediction for $\dg$ is based on the fact that the decay width into
CP-even final states is larger than into CP-odd final states.
When new CP violating phases appear in the mixing amplitude, then
the mass eigenstates can differ significantly from the CP eigenstates,
and both mass eigenstates are allowed to decay into the
CP-even final states. Consequently, $\dg$ is reduced.

\newsec{Extensions of the Scalar Sector}

\subsec{Charged Scalar Exchange}
When Natural Flavor Conservation (NFC) is maintained
CP violation could arise in charged scalar exchange if there were
at least three Higgs doublets
\ref\WeinCS{S. Weinberg, Phys. Rev. Lett. 37 (1976) 657.}.
This is also the minimal number of doublets required when CP breaking
is spontaneous only
\ref\BranNFC{G.C. Branco, Phys. Rev. Lett. 44 (1980) 504.}.
In this case, $\delta_{\rm KM}=0$ and all CP violation comes from the
mixing of scalar fields. This model is ruled out, as we show below.
It is, of course, still a viable possibility that CP is
explicitly broken, in which case both quark and Higgs mixings
provide CP violation.

We investigate a multi Higgs doublet model (with $n\ge3$ doublets) with
NFC and assume that it is a different scalar
that couples to the down, up and lepton sectors:
\eqn\LCSi{-\L_Y=
-{\phi_1^+\over v_1}\overline{U}VM_d^{\rm diag}P_R \, D
+{\phi_2^+\over v_2}\overline{U}M_u^{\rm diag}V P_L \,D
-{\phi_3^+\over v_3}\overline{\nu}M_\ell P_R \, \ell + {\rm h.c.},}
where $V$ is the CKM matrix and $P_{L,R}=(1\mp\gamma_5)/2$.
We denote the physical charged scalars by $H_i^+$ ($i=1,2,\ldots,n-1$),
and the would-be Goldstone boson (eaten by the $W^+$) by $H_n^+$. We
define $K$ to be the matrix that rotates the charged scalars from the
interaction- to the mass-eigenbasis. Then the Yukawa Lagrangian in the
mass basis (for both fermions and scalars) is
\eqn\LCSm{\L_Y={G_F^{1/2}\over2^{1/4}}\sum_{i=1}^{n-1}
\{H_i^+\overline{U}[Y_i M_u^{\rm diag}V P_L +X_i VM_d^{\rm diag}P_R ]D  +
H_i^+\overline{\nu} [Z_i M_\ell P_R ]\ell  \}+{\rm h.c.},}
where
\eqn\XYdef{X_i=-{K_{i1}^*\over K_{n1}^*},\ \ \
Y_i=-{K_{i2}^*\over K_{n2}^*},\ \ \
Z_i=-{K_{i3}^*\over K_{n3}^*}.}
CP violation in the charged scalar sector comes from phases in the mixing
matrix for charged scalars. CP violating effects are largest when
the lightest charged scalar is much lighter than the heaviest one
\nref\WeiL{S. Weinberg, Phys. Rev. D42 (1990) 860.}%
\nref\Lav{L. Lavoura, Int. J. Mod. Phys. A8 (1993) 375.}%
\refs{\WeiL,\Lav}. Here we assume that all but the lightest charged
scalar ($H_1^+$) effectively decouple from the fermions.
Then, CP violating observables depend on three parameters:
\eqn\threeCPV{\eqalign{
{\Im(XY^*)\over m_H^2} \equiv\ &\ {\Im(X_1Y_1^*) \over m_{H_1}^2}\
\approx\ \sum_{i=1}^{n-1} {\Im(X_iY_i^*) \over m_{H_i}^2},\cr
{\Im(XZ^*)\over m_H^2} \equiv\ &\ {\Im(X_1Z_1^*) \over m_{H_1}^2}\
\approx\ \sum_{i=1}^{n-1} {\Im(X_iZ_i^*) \over m_{H_i}^2},\cr
{\Im(YZ^*)\over m_H^2} \equiv\ &\ {\Im(Y_1Z_1^*) \over m_{H_1}^2}\
\approx\ \sum_{i=1}^{n-1} {\Im(Y_iZ_i^*) \over m_{H_i}^2}.\cr }}
$\Im(XY^*)$ induces CP violation in the quarks sector, while $\Im(XZ^*)$
and $\Im(YZ^*)$ give CP violation that is observable in
semi-leptonic processes.

As mentioned above, there is an interesting question of whether charged
scalar exchange could be the {\it only} source of CP violation. In other
words, we would like to know whether a model of extended scalar sector
with spontaneous CP violation and NFC is viable. It is not clear that
the model could account for $\epsK$
\nref\Sand{A.I. Sanda, Phys. Rev. D23 (1981) 2647.}%
\nref\Desh{N.G. Deshpande, Phys. Rev. D23 (1981) 2654.}%
\nref\DoHo{J.F. Donoghue and B.R. Holstein, Phys. Rev. D32 (1985) 1152.}%
\nref\Chen{H.Y. Cheng, Phys. Rev. D34 (1986) 1397.}%
\nref\BiSa{I.I. Bigi and A.I. Sanda, Phys. Rev. Lett. 58 (1987) 1605.}%
\nref\Che{H.Y. Cheng, Phys. Rev. D42 (1990) 2329.}%
\refs{\Sand-\Che}. But if it does, then the charged scalar contribution
to $d_N$
\ref\KrPo{P. Krawczyk and S. Pokorski, Nucl. Phys. B364 (1991) 11.}\
and, more convincingly, to $\Gamma(b\ra s\gamma)$
\ref\GrNis{Y. Grossman and Y. Nir, Phys. Lett. B313 (1993) 126.}\
are too large. We now explain this point in more detail.

In this framework, neither short distance contributions nor long
distance ones from intermediate $2\pi$ state can produce large enough
$\epsK$. One needs to assume that the
dominant contribution comes from an intermediate $\eta_0$:
\eqn\dometa{\epsK\approx{e^{i\pi/4}\over\sqrt2\Delta m_K}\Im{
\vev{K^0|H|\eta_0}\vev{\eta_0|H|\bar K^0}\over m_K-m_{\eta_0}}.}
To account for the numerical value of $\epsK$, the charged scalar
parameters should fulfill \refs{\BiSa-\Che}
\eqn\epsbou{{\Im(XY^*)\over m_H^2}
\left[\ln{m_H^2\over m_c^2}-{3\over2}\right]=0.11\ GeV^{-2}.}
With $m_{H}\geq42\ GeV$, this gives
\eqn\epslim{\Im(XY^*)\gsim40.}

A large contribution to the EDM of the neutron
$d_N$ comes from the EDM of the down quark \KrPo:
\eqn\dNCS{d_N^{(d)}={\sqrt2 G_F m_d\over9\pi^2}\Im(XY^*)\left[
\eta_c|V_{cd}|^2g(m_c^2/m_H^2)+\eta_t|V_{td}|^2g(m_t^2/m_H^2)\right],}
with
\eqn\gfun{g(x)={x\over(1-x)^2}\left[{5x\over4}-{3\over4}-
{1-3x/2\over1-x}\ln x\right].}
With some conservative assumptions, and using the lower bound \epslim,
we get $d_N\gsim2.5\times10^{-25}\ e\ {\rm cm}$, a factor of 2 above
the experimental upper bound. An even larger contribution comes
through the three gluon operator
\ref\Dicu{D. Dicus, Phys. Rev. D41 (1990) 999.}:
\eqn\dNCSg{\eqalign{d_N^{(g)}= 4\times10^{-21}\ e\ {\rm cm}
\left({\alpha_s(m_W)\over\alpha_s(m_b)}\right)^{14/23}&
\left({\alpha_s(m_b)\over\alpha_s(m_c)}\right)^{54/25}
\left({\alpha_s(m_c)\over\alpha_s(m_N)}\right)^{54/27}\cr \times&
\left({g_s(m_N)\over4\pi}\right)^3\
\Im(XY^*)F_3\left({m_t^2\over m_H^2}\right),\cr}}
where
\eqn\Fthfun{F_3(x)={x\over4(1-x)^3}(-3+4x-x^2-2\ln x).}
This result, which suffers from large hadronic uncertainties,
seems to be about two orders of magnitude above the bound.

The strongest constraint on $\Im(XY^*)$, however, comes -- somewhat
surprisingly -- from a CP conserving process, the decay $b\ra s\gamma$
\ref\CLEO{M.S. Alam {\it et al.}, CLEO Collaboration,
 Phys. Rev. Lett. 74 (1995) 2885.}:
\eqn\bsgexp{{\rm BR}(b\ra s\gamma)\leq4.2\times10^{-4}.}
Within our model, this ratio is given by \KrPo:
\eqn\bsgthe{{\rm BR}(b\ra s\gamma)=C\left|\eta_2+G_W(x_t)+
(|Y|^2/3)G_W(y_t)+(XY^*)G_H(y_t)\right|^2,}
where
\eqn\Ccon{C\equiv{3\alpha\eta_1^2{\rm BR}(B\ra X_c\ell\nu)\over2\pi
F_{\rm ps}(m_c^2/m_b^2)}\approx3\times10^{-4},}
$F_{\rm ps}\sim0.5$ is a phase space factor, $\eta_1\sim0.66$ and
$\eta_2\sim0.57$ are QCD correction factors
\ref\GSW{B. Grinstein, R. Springer and M. Wise, Nucl. Phys. B339
(1990) 269.},
$x_t=m_t^2/m_W^2$, $y_t=m_t^2/m_H^2$, and the expressions
for the $G$ functions can be found in \KrPo.

The upper bound on $\Im(XY^*)$ corresponds to a case where the
real part of the new diagrams cancels the Standard Model contributions
and the upper bound \bsgexp\ is saturated by the imaginary part
of these diagrams \GrNis:
\eqn\bsgbou{\Im(XY^*)\lsim\sqrt{4.2\times10^{-4}\over C}{1\over
G_H(y_t)}.}
For $m_t\sim180\ GeV$, we get
\eqn\bsgboua{\Im(XY^*)\lsim\cases{1.6&$m_H\sim{1\over2}m_Z$,\cr
3.2&$m_H\sim2m_Z$.\cr}}

The upper bound on $\Im(XY^*)$ implies that charged scalar exchange
can make only a negligible contribution to $\epsK$ and cannot
be the only source of CP violation \GrNis. A detailed investigation shows
that, in spite of the fact that charged scalar contributions could, in
principle, contribute to $B-\bar B$ mixing with new phases, this
contribution is numerically small and would modify the Standard
Model predictions for CP asymmetries in $B$ decays by no more
than $\O(0.02)$ \GrNis. On the other hand, the contribution
to $d_N$ can still be close to the experimental upper bound.

\subsec{Transverse Lepton Polarization}
As triple-vector correlation is odd under time-reversal, the experimental
observation of such correlation would signal T and -- assuming CPT
symmetry -- CP violation.\foot{It is possible to get non-vanishing
T$-$odd observables even without CP violation (see e.g.
\ref\FSIphase{M.B. Gavela {\it et al.}, Phys. Rev. D39 (1989) 1870}).
Such ``fake" asymmetries can arise due to CP conserving unitary phases
from final state interactions (FSI).
They can be removed by comparing the measurements in two CP conjugate
channels.} The muon transverse polarization in $K\to\pi\,\mu\,\nu$ decays
\nref\LeWU{T.D. Lee and C.S. Wu, Ann. Rev. Nucl. Sci. 16 (1966) 471.}%
\nref\Che{H.-Y. Cheng, Phys. Rev. D28 (1983) 150.}%
\nref\Miriam{M. Leurer, Phys. Rev. Lett. 62 (1989) 1967.}%
\nref\CFK{P. Castoldi, J.M. Fr\`ere and G. Kane,
 Phys. Rev. D39 (1989) 2633.}%
\nref\GaKa{R. Garisto and G. Kane, Phys. Rev. D44 (1991) 2038.}%
\nref\BeGe{G. B\'elanger and C.Q. Geng,  Phys. Rev. D44 (1991) 2789.}%
\refs{\LeWU-\BeGe}, and the tau transverse
polarization in semileptonic heavy quark decays
\nref\allheavy{H.-Y. Cheng, Phys. Rev. D26 (1982) 143.}%
\nref\GolVal{E. Golowich and G. Valencia, Phys. Rev. D40 (1989) 112.}%
\nref\Eilam{D. Atwood, G. Eilam and A. Soni,
 Phys. Rev. Lett. 71 (1993) 492.}%
\nref\Garisto{R. Garisto, Phys. Rev. D51 (1995) 1107.}%
\refs{\allheavy-\Garisto} are examples of such observables.
The lepton transverse polarization cannot be generated by vector or
axial-vector interactions only \refs{\Miriam,\CFK}, so it is particularly
suited for searching for CP violating scalar contributions.

The lepton transverse polarization, $P_\perp$, in semileptonic decays
is defined as the lepton polarization component along the normal vector
of the decay plane,
\eqn\defpper{P_\perp = {\vec s_{\ell} \cdot (\vec p_{\ell} \times
\vec p_X) \over |\vec p_{\ell} \times \vec p_X|}\,,}
where $\vec s_{\ell}$ is the lepton spin three-vector and $\vec p_{\ell}$
($\vec p_X$) is the three-momentum of the lepton (hadron).
Experimentally, it is useful to define the integrated CP violating
asymmetry
\eqn\defacp{a_{CP} \equiv \langle P_\perp \rangle =
{\Gamma^+ - \Gamma^- \over \Gamma^+ + \Gamma^-},}
where $\Gamma^+$ ($\Gamma^-$) is the rate of finding the lepton spin
parallel (anti-parallel) to the normal vector of the decay plane.

A measurable non-zero $a_{CP}$ will be a clear signal of new sources of
CP violation beyond the SM. The SM predictions and the ``fake"
asymmetries are much smaller than the current experimental sensitivity
\nref\KfsiGS{E.S. Ginsberg and J. Smith, Phys. Rev. D8 (1973) 3887.}%
\nref\kpfsi{A.R. Zhitnitskii, Yad. Fiz. 31 (1980) 1024
[Sov. J. Nucl. Phys. 31 (1980) 529].}%
\nref\KfsiAd{G.S. Adkins, Phys. Rev. D28 (1983) 2885.}%
\nref\KSW{J.G. K\"orner, K. Schilcher and Y.L. Wu,
 Phys. Lett. B242 (1990) 119.}%
\refs{\KfsiGS-\KSW}. A non-zero $a_{CP}$ can arise in our model
from the interference between the $W$-mediated and the $H^+$-mediated
tree diagrams. For strange and bottom quark decays,
the asymmetry is given by
\eqn\acpdown{a_{CP} = C_{ps} {\Im(XZ^*) \over m_H^2}, }
while for charm and top quark decays, it is given by
\eqn\acpup{a_{CP} = C_{ps} {\Im(YZ^*)\over m_H^2} .}
The $C_{ps}$ factor is different for each decay mode. It depends
on the phase space integrals, the form factors and masses involved,
and has even a mild dependence on $XZ^*$ and $YZ^*$ as they affect
the total decay rates.

To find how large can these asymmetries be, we study the bounds on the
CP violating parameters. In the down sector, the strongest bound is
obtained from the measurement of the inclusive $b\to X\tau\nu$ decay
\ref\pdg{R.M. Barnett {\it et al.}, Phys. Rev. D54 (1996) 1.}.
At the $2 \sigma$ level it reads
\nref\KrPobc{P. Krawczyk and S. Pokorski,
 Phys. Rev. Lett. 60 (1988) 182.}%
\nref\Kali{J. Kalinowski, Phys. Lett. B245 (1990) 201.}%
\nref\GrLi{Y. Grossman and Z. Ligeti, Phys. Lett. B332 (1994) 373.}%
\nref\GrHaNi{Y. Grossman, H.E. Haber and Y. Nir,
 Phys. Lett. B357 (1995) 630.}%
\refs{\KrPobc-\GrHaNi}:
\eqn\boundrim{{\Im(XZ^*) \over m_H^2} < 0.16 \, GeV^{-2}.}

For the $K^+ \to \pi^0 \mu^+ \nu$ decay, \boundrim\ implies
\nref\Kuno{Y. Kuno, Nucl. Phys. B (Proc. Suppl.) 37A (1994) 87.}
\refs{\Garisto,\Kuno}
\eqn\acpKal{a_{CP}(K^+\to\pi^0\mu^+\nu)\lsim 8\times10^{-3},}
which is close to the current experimental bound \pdg\
${a_{CP} (K^+ \to \pi^0 \mu^+ \nu) < 1.2 \times 10^{-2}}$.
Since scalars couple more strongly to heavier fermions, the expected
signals are stronger in heavy quark decays.
For inclusive $B$ decays, \boundrim\ implies
\nref\GrLicp{Y. Grossman and Z. Ligeti, Phys. Lett. B347 (1995) 399.}%
\refs{\Eilam,\GrLicp}
\eqn\acpb{a_{CP}(B \to X \tau \nu) \lsim 0.3.}
For exclusive $B$ decays \Garisto, the asymmetries are larger but
the theoretical uncertainties are also larger.
The allowed asymmetries for decays into muon are suppressed by
$m_\mu / m_\tau$ and, in addition, the muon spin is harder to tag.

In the up sector, the experimental bound on $\Im(YZ^*)$ is just the
product of the bounds on $Y$ and $Z$
\ref\GrMHDM{Y. Grossman, Nucl. Phys. B426 (1994) 355.}.
The strongest bound on $|Y|$ comes from the measurement of
$R_b \equiv {Z \to b \bar b \over Z \to {\rm hadrons}}$.
Requiring that the charged Higgs contribution to $R_b$ does not exceed
$0.003$, we get
\eqn\boundY{|Y| \lsim\cases{1.2&$m_H\sim{1\over2}m_Z$,\cr
1.6&$m_H\sim2m_Z$.\cr}}
The charged Higgs contribution is proportional to $|Y|^2$,
and the full dependence on $m_H$ can be found in
\ref\BoFi{M. Boulware and D. Finnell, Phys. Rev. D44 (1991) 2054.}.
The bound on $|Z|$ comes from lepton universality in tau decay:
\eqn\boundZ{{|Z| \over m_H} \leq 1.7 \, {\rm GeV^{-1}} }

For exclusive $D$ decays, \boundY\ and \boundZ\ imply that the allowed
asymmetries are $\lsim\O(10^{-2})$ \Garisto. For top decays, we get
\eqn\acptb{a_{CP}( t \to b \tau \nu) \lsim 4\times10^{-2}.}
Choosing an optimal part of phase space can enhance the signal by a
factor of about $5$ \Eilam. Since the $W$ is on shell, several other
observables can be constructed for top decays \Eilam.

To conclude: Multi Higgs doublet models can give a measurable signal for
transverse lepton polarization in $K$, $B$ and top decays. Such a signal
is a clear indication of New Physics.

\subsec{Flavor Changing Neutral Scalar Exchange}

Natural flavor conservation needs not be exact in models of extended
scalar sector
\nref\BrRe{G.C. Branco and M.N. Rebelo, Phys. Lett. B160 (1985) 117.}%
\nref\LiWo{J. Liu and L. Wolfenstein,  Nucl. Phys. B289 (1987) 1.}%
\nref\LiWoB{J. Liu and L. Wolfenstein, Phys. Lett. B197 (1987) 536.}%
\nref\HaNiS{H. Haber and Y. Nir, Nucl. Phys. B335 (1990) 363.}%
\refs{\BrRe-\HaNiS}. In particular, it is quite likely that the existence
of the additional scalars is related to flavor symmetries that explain
the smallness and hierarchy in the Yukawa couplings. In this case, the
new flavor changing couplings of these scalars are suppressed by the
same selection rules as those that are responsible to the smallness
of fermion masses and mixing, and there is no need to impose NFC
\nref\ChSh{T.P. Cheng and M. Sher, Phys. Rev. D35 (1987) 3484.}%
\nref\JoRi{A.S. Joshipura and S.D. Rindani,
 Phys. Lett. B260 (1991) 149.}%
\nref\AHR{A. Antaramian, L.J. Hall and A. Rasin,
 Phys. Rev. Lett. 69 (1992) 1871.}%
\nref\WoWu{Y.L. Wu and L. Wolfenstein, Phys. Rev. Lett. 73 (1994) 1762.}%
\nref\BGL{G.C. Branco, W. Grimus and L. Lavoura,
 Phys. Lett. B380 (1996) 119.}%
\nref\ARS{D. Atwood, L. Reina and A. Soni,
 Phys. Rev. D54 (1996) 3269.}%
\refs{\HaWeS,\ChSh-\ARS}. An explicit framework, with Abelian horizontal
symmetries, was presented in
\nref\LNS{M. Leurer, Y. Nir and N. Seiberg,
 Nucl. Phys. B398 (1993) 319.}%
\refs{\LNS,\LNSb}. (For another related study, see
\ref\ACHM{N. Arkani-Hamed, C.D. Carone, L.J. Hall and H. Murayama,
 Phys. Rev. D54 (1996) 7032.}.)
We explain the general idea using these models.

The simplest model of ref. \LNS\ extends the SM by supersymmetry and by
an Abelian horizontal symmetry $\H=U(1)$ (or $Z_N$). The symmetry $\H$ is
broken by a VEV of a single scalar $S$ (to which we attribute charge
$\H(S)=-1$) that is a singlet of the SM gauge group. Consequently, Yukawa
couplings that violate $H$ arise only from nonrenormalizable terms.
Defining the relevant sums of $\H$ charges through
\eqn\defnij{\eqalign{n^d_{ij}=\H(Q_i)+\H(\bar d_j)+\H(\phi_d),\cr
n^u_{ij}=\H(Q_i)+\H(\bar u_j)+\H(\phi_u),\cr}}
we find the following form for scalar-fermion couplings:
\eqn\YukH{
\L_Y=X^d_{ij}\left({S\over M}\right)^{n^d_{ij}} Q_i\bar d_j\phi_d
+X^u_{ij}\left({S\over M}\right)^{n^u_{ij}}Q_i\bar u_j\phi_u,}
where $M$ is some high energy scale and $X^d_{ij},X^u_{ij}$ are $\O(1)$
complex numbers.\foot{In the supersymmetric framework,
$n^q_{ij}<0$ implies $X^q_{ij}=0$ due to the holomorphy of the
superpotential.} For a singlet VEV $\vev{S}\ll M$, a small parameter
$\lambda=\vev{S}/M$ suppresses $\H$ violating terms. More precisely,
the effective Yukawa terms,
\eqn\Yukeff{\L_Y^{\rm eff}=
Y^d_{ij}Q_i\bar d_j\phi_d+Y^u_{ij}Q_i\bar u_j\phi_u,}
obey selection rules that can be read from \YukH\ ($q=u,d$):
\eqn\select{Y^q_{ij}=\O\left[\lambda^{n^q_{ij}}\right].}
Note that each Yukawa coupling is proportional, in addition to the
suppression factors \select, to a {\it complex} coefficient of order 1.

The smallness and hierarchy in the quark (and lepton) masses and
mixing arises now in a natural way. Explicitly (for $i<j$):
\eqn\masmix{\eqalign{|V_{ij}|\sim&\ \lambda^{\H(Q_i)-\H(Q_j)},\cr
m(d_i)/m(d_j)\sim&\
\lambda^{\H(Q_i)-\H(Q_j)+\H(\bar d_i)-\H(\bar d_j)},\cr
m(u_i)/m(u_j)\sim&\
\lambda^{\H(Q_i)-\H(Q_j)+\H(\bar u_i)-\H(\bar u_j)}.\cr}}
If, for example, we take $\lambda\sim0.2$, then the order of magnitude
of the three CKM mixing angles and of the six quark masses
arise naturally for the following $\H$ charge assignments:
\eqn\Hcharge{\matrix{Q_1&Q_2&Q_3&&\bar d_1&\bar d_2&\bar d_3\cr
(3)&(2)&(0)&&(3)&(2)&(2)\cr
\bar u_1&\bar u_2&\bar u_3&&&\phi_d&\phi_u\cr
(3)&(1)&(0)&&&(0)&(0)\cr}}
(where we took $\tan\beta=\vev{\phi_u}/\vev{\phi_d}\sim1$).

The singlet scalar $S$ has flavor changing couplings, $Z^q_{ij}$
($q=u,d$; $i,j=1,2,3$). Their magnitude is related to that of the
effective Yukawa couplings $Y^q_{ij}$:
\eqn\relZY{Z^q_{ij}\sim {\vev{\phi_q}\over\vev{S}}\ Y^q_{ij}.}
These couplings contribute, for example, to $K-\bar K$
mixing proportionally to
\eqn\FCNS{Z^d_{12}Z^{d*}_{21}\sim{m_d m_s\over\vev{S}^2},}
where we used \select, \masmix\ and \relZY\ to estimate the
magnitude of the flavor changing couplings in terms of known
quark parameters.

With arbitrary phase factors in the various $Z^q_{ij}$ couplings,
the contributions to neutral meson mixing are, in general, CP violating.
In particular, there will be a contribution to $\epsK$ from
$\Im(Z^d_{12}Z^{d*}_{21})$. Requiring that the $S$-mediated tree level
contribution does not exceed the experimental value of $\epsK$ gives,
for $\O(1)$ phases,
\eqn\bouFCNS{M_S\vev{S}\gsim1.8\ TeV^2.}
We learn that (for $M_S\sim\vev{S}$) the mass of the $S$-scalar
could be as low as 1.5\ TeV, some 4 orders of magnitude below the
bound corresponding to $\O(1)$ flavor changing couplings.

The flavor changing couplings of the $S$-scalar lead also to
a tree level contribution to $B-\bar B$ mixing proportional to
\eqn\FCNSB{Z^d_{13}Z^{d*}_{31}\sim{m_d m_b\over\vev{S}^2}.}
This means that, for phases of order 1, the neutral scalar exchange
accounts for at most a few percent of $B-\bar B$ mixing. This cannot
be signaled in $\Delta m_B$ (because of the hadronic uncertainties
in the calculation) but could be signalled (if $\vev{S}$ is at the
lower bound) in CP asymmetries in $B^0$ decays.

Finally, the contribution to $D-\bar D$ mixing, proportional to
\eqn\FCNSC{Z^u_{12}Z^{u*}_{21}\sim{m_u m_c\over\vev{S}^2},}
is below a percent of the current experimental bound. This is unlikely
to be discovered in near-future experiments, even if the new phases
maximize the interference effects in the $D^0\ra K^-\pi^+$ decay.

To summarize, models with horizontal symmetries naturally suppress
flavor changing couplings of extra scalars. There is no need to
invoke NFC even for new scalars at the TeV scale. Furthermore,
the magnitude of the flavor changing couplings is related to the
observed fermion parameters. Typically, contributions from
neutral scalars with flavor changing couplings could dominate
$\epsK$. If they do, then a signal at the few percent level
in CP asymmetries in neutral $B$ decays is quite likely.

\subsec{Neutral Scalar Exchange in Top Physics}

It is possible that the neutral scalars are mixtures
of CP-even and CP-odd scalar fields
\nref\TDLee{T.D. Lee, Phys. Rev. D8 (1973) 1226.}%
\nref\AST{C.H. Albright, J. Smith, and S.H.H. Tye,
 Phys. Rev D21 (1980) 711.}%
\nref\MePo{A. M\'endez and A. Pomarol, Phys. Lett. B272 (1991) 313.}%
\nref\PoVe{A. Pomarol and R. Vega, Nucl. Phys. B413 (1993) 3.}%
\refs{\TDLee-\PoVe,\WeiL-\Lav,\BrRe}. Such a scalar couples to
both scalar and pseudoscalar currents:
\eqn\Lyuxmix{\L_Y=H_i\bar f(a_i^f+ib_i^f\gamma_5)f,}
where $H_i$ is the physical Higgs boson and $a_i^f,b_i^f$ are functions
of mixing angles in the matrix that diagonalizes the neutral scalar
mass matrix. (Specifically, they are proportional to the components
of, respectively, $\Re\phi_u$ and $\Im\phi_u$ in $H_i$.)
CP violation in processes involving fermions is
proportional to $a_i^fb_i^{f*}$. The natural place to look for
manifestations of this type of CP violation is top physics,
where the large Yukawa couplings allow large asymmetries
\nref\Pes{C.E. Schmidt and M.E. Peskin, Phys. Rev. Lett. 69 (1992) 410.}%
\nref\Cha{D. Chang and W.Y. Keung, Phys. Lett. B305 (1993) 261.}%
\nref\Pil{A. Pilaftsis and M. Nowakowski,
 Int. J. Mod. Phys. A9 (1994) 1097, (E) A9 (1994) 5849.}%
\nref\HeMa{X.G. He, J.P. Ma and B. McKellar,
 Mod. Phys. Lett. A9 (1994) 205.}%
\nref\BeBr{W. Bernreuther and A. Brandenburg,
 Phys. Lett. B314 (1993) 104.}%
\nref\ABB{H. Anlauf, W. Bernreuther and A. Brandenburg,
 Phys. Rev. D52 (1995) 3803, (E) D53 (1996) 1725.}%
\nref\Grza{B. Grzadkowski,  Phys. Lett. B338 (1994) 71.}%
\nref\AGS{T. Arens, U.D.J. Gieseler and L.M. Sehgal,
 Phys. Lett. B339 (1994) 127.}%
\nref\BAEM{S. Bar-Shalom, D. Atwood, G. Eilam, R.R. Mendel and A. Soni,
 Phys. Rev. D53 (1996) 1162.}%
\refs{\Pes-\BAEM}. Note that unlike our discussion above, the asymmetries
here have nothing to do with FCNC processes. Actually, in models with
NFC (even if softly broken \BrRe), the effects discussed here contribute
negligibly to $\epsK$ and to CP asymmetries in $B$ decays. On the
other hand, two loop diagrams with intermediate neutral scalar and top
quark can induce a CP violating three gluon operator
\refs{\Weintg,\Dicu} that would give $d_N$ close to the experimental
bound
\nref\GuWy{J.F. Gunion and D. Wyler, Phys. Lett. B248 (1990) 170.}%
\nref\BLY{E. Braaten, C.S. Li and T.C. Yuan,
 Phys. Rev. Lett. 64 (1990) 1709.}%
\nref\RGPV{A. De R\'ujula, M.B. Gavela, O. P\`ene and F.J. Vegas,
 Phys. Lett. B245 (1990) 640.}%
\refs{\Dicu,\GuWy-\RGPV}.

\subsec{The Superweak Scenario}

CP violation via neutral scalar exchange is the most commonly
studied realization of the {\it superweak} scenario
\ref\SuperWeak{L. Wolfenstein, Phys. Rev. Lett. 13 (1964) 562.}.
The original scenario stated that CP violation appears in a new
$\Delta S=2$ interaction while there is no CP violation in the SM
$\Delta S=1$ transitions. Consequently, the only large observable
CP violating effect is $\epsK$, while $\epe\sim10^{-8}$ and EDMs are
negligibly small. At present, the idea of ``superweak CP violation"
refers to many different types of models. There are several reasons for
this situation:

(i) The work of ref. \SuperWeak\ was concerned only with
CP violation in $K$ decays. In extending the idea to other mesons,
one may interpret the idea in various ways. On one side, it is possible
that the superweak interaction is significant only in $K-\bar K$
mixing and (apart from the relaxation of the $\epsK$-bounds on the
CKM parameters) has no effects on mixing of heavier mesons.
On the other extreme, one may take the superweak scenario to imply
that CP violation comes from $\Delta F=2$ processes only for all mesons.

(ii) The scenario proposed in \SuperWeak\ did not employ any specific
model. It was actually proposed even before the formulation of the
Standard Model. To extend the idea to, for example, the neutral $B$
system, a model is required. Various models give very different
predictions for CP asymmetries in $B$ decays.

(iii) It is rather difficult to achieve the superweak scenario
in a natural way. In particular, it is difficult to understand why
would CP be a good symmetry in one sector of the theory but not
in another. Consequently, in most models that employ approximate
symmetries, the CKM phase does not vanish and the resulting CP violation
is quite different from the original scenario.

(iv) The question of whether CP violation can occur in $\Delta F=2$
transitions only is not always well-defined. When discussing CP violation
in the interference of mixing and decay, it is a matter of convention
to decide whether to put the CP violating phases in $q/p$ ($\Delta F=2$)
or $\bar A/A$ ($\Delta F=1$) or both. The common use of the term
`superweak CP violation' refers to a situation where $|\bar A/A|=1$
and there exists a convention where $\bar A/A=1$ for all processes.

If one extends the superweak scenario to the $B$ system by assuming that
there is CP violation in $\Delta b=2$ but not in $\Delta b=1$
transitions, the prediction for CP asymmetries in $B$ decays into final
CP eigenstates is that they are equal for all final states
\nref\Wins{B. Winstein, Phys. Rev. Lett. 68 (1992) 1271.}%
\nref\SoWo{J.M. Soares and L. Wolfenstein, Phys. Rev. D46 (1992) 256.}%
\nref\WiWo{B. Winstein and L. Wolfenstein,
 Rev. Mod. Phys. 65 (1993) 1113.}%
\refs{\Wins-\WiWo}. Whether these asymmetries are all small or could be
large is model dependent. In addition, the asymmetries in charged $B$
decays vanish.

Various models (or scenarios) that realize the main features of the
superweak idea can be found in refs.
\nref\GeNa{J.-M. Gerard and T. Nakada, Phys. Lett. B261 (1991) 474.}%
\nref\LavSW{L. Lavoura, Int. J. Mod. Phys. A9 (1994) 1873.}%
\refs{\GeNa-\LavSW,\LiWo-\LiWoB}. As mentioned above, there is
a considerable variation in their predictions for $\epe$, $d_N$ and
other quantities. Note, in particular, that neither a measurement of
$\epe$ at the level of $10^{-4}$ nor of $d_N$ at the level of $10^{-26}\
e$ cm will unambiguously exclude these models.

\newsec{Left Right Symmetry}

\subsec{The Theoretical Framework}

We study a specific version of Left-Right Symmetric (LRS) models, where
P, C and CP are symmetries of the Lagrangian that are spontaneously
broken
\nref\Chan{D. Chang, Nucl. Phys. B214 (1983) 435.}%
\nref\BFG{G.C. Branco, J.-M. Fr\`ere and J.-M. G\'erard,
 Nucl. Phys. B221 (1983) 317.}%
\nref\HaLe{H. Harari and M. Leurer, Nucl. Phys. B233 (1984) 221.}%
\nref\EGN{G. Ecker, W. Grimus and H. Neufeld,
 Nucl. Phys. B247 (1983) 70.}%
\nref\EGN{G. Ecker and W. Grimus, Nucl. Phys. B258 (1985) 328.}%
\nref\Leur{M. Leurer, Nucl. Phys. B266 (1986) 147.}%
\refs{\Chan-\Leur}. The electroweak gauge group is $SU(2)_L\times
SU(2)_R\times U(1)_{B-L}$. Left-handed quarks reside in $Q_L(2,1)_{1/3}$
representations and right-handed ones in $Q_R(1,2)_{1/3}$. The scalar
content of the minimal LRS models is
\ref\MoPa{R.N. Mohapatra and J.C. Pati, Phys. Rev. D11 (1975) 566.}\
$\Phi(2,2)_0$, $\Delta_L(3,1)_2$ and $\Delta_R(1,3)_2$.
A model with only minimal scalar sector and spontaneous CP violation
predicts unacceptably large FCNC
\ref\BLMW{J. Basecq, J. Liu, J. Milutonovic and L. Wolfenstein,
 Nucl. Phys. B272 (1986) 145.}.
To avoid this, one has to add scalar singlets or triplets but these
do not affect our analysis. The only specific assumption about the
scalar sector that we make is the existence of a single $\Phi$ field.
The VEV of $\Phi$ is
\eqn\vevPhi{\vev{\Phi}=\pmatrix{k&0\cr0&k^\prime e^{i\eta}\cr}.}
The relative phase between $k$ and $k^\prime$, $\eta$, spontaneously
breaks CP. In principle, it is the only source of CP violation.
Eventually, there are seven CP violating phases in the mass eigenbasis.
They all vanish when $\eta=0$, but practically they are independent
parameters.

The phase $\eta$ appears explicitly in the mixing of the charged
gauge bosons:
\eqn\mixWLR{\eqalign{W_1=&\cos\xi\ W_L+e^{-i\eta}\sin\xi\ W_R,\cr
W_2=&-e^{i\eta}\sin\xi\ W_L+\cos\xi\ W_R,\cr}}
where
\eqn\defxi{\xi={k k^\prime\over\vev{\Delta_R^0}^2}.}
The Yukawa couplings are given by
\eqn\Yuklrs{\L_Y=\overline{Q_L}(A\Phi+B\tau_2\Phi^*\tau_2)Q_R
+{\rm h.c.},}
where $\tau_2$ is the Pauli matrix acting in the $SU(2)_L$ or $SU(2)_R$
space, $A$ and $B$ are matrices in generation space.

P symmetry requires that $A$ and $B$ are hermitian;
C symmetry requires that $A$ and $B$ are symmetric;
and CP invariance implies that $A$ and $B$ are real. The mass matrices,
\eqn\masmat{\eqalign{M_u\ =&\ kA+k^\prime e^{-i\eta}B,\cr
M_d\ =&\ k^\prime e^{i\eta}A+kB,\cr}}
are symmetric. The symmetry of the matrices implies that
\eqn\relVF{V_R=F_u V_L^* F_d^\dagger,}
where $V_L$ and $V_R$ are the charged current mixing matrices for
$W_L$ and $W_R$, respectively, while $F_u$ and $F_d$ are diagonal
unitary matrices:
\eqn\defF{F_u={\rm diag}(e^{i\phi_u},e^{i\phi_c},e^{i\phi_t});\ \ \
F_d={\rm diag}(e^{i\phi_d},e^{i\phi_s},e^{i\phi_b}).}
On top of the single CP violating phase of the CKM matrix $V_L$,
there are 5 phase differences in $F_u,F_d$.

For the purpose of studying new contributions to CP violation,
it is simpler to work in a two generation framework. In this case,
$V_L$ is real and there are 3 phases in $F_u,F_d$. We define:
\eqn\defLRS{\eqalign{
\gamma=&(\phi_c+\phi_u-\phi_s-\phi_d)/2+\eta,\cr
\delta_1=&(\phi_c-\phi_u+\phi_s-\phi_d)/2,\cr
\delta_2=&(\phi_c-\phi_u-\phi_s+\phi_d)/2.\cr}}
Choosing a basis where $V_L$ is real and the mixing of $W_L-W_R$ is real,
these phases appear in $V_R$ only:
\eqn\mixLRS{V_W=\pmatrix{c_\xi&s_\xi\cr-s_\xi&c_\xi\cr},\ \
V_L=\pmatrix{c_\theta&s_\theta\cr-s_\theta&c_\theta\cr},\ \
V_R=e^{i\gamma}\pmatrix{e^{-i\delta_2}c_\theta&
e^{-i\delta_1}s_\theta\cr-e^{i\delta_1}s_\theta&
e^{i\delta_2}c_\theta\cr}.}

\subsec{Phenomenological Consequences}

For $\epsK$, the dominant contribution comes from box diagrams
with both $W_L$ and $W_R$ in the loop and from tree level diagrams
mediated by the extra Higgs doublet. $W_L-W_R$ mixing can be safely
neglected. The value of $M_{12}(K)$ in this model is \refs{\HaLe,\Leur}
\eqn\KmixLRS{{M_{12}^{\rm LRS}\over M_{12}^{\rm SM}}=
1-e^{i(\delta_2-\delta_1)}\left[430\beta-15\beta\ln\beta+Q_H^2
(11600\beta_H-15\beta_H\ln\beta_H)\right],}
where
\eqn\defbetaH{\beta={m^2_{W_1}\over m^2_{W_2}},\ \ \
\beta_H={m^2_{W_1}\over m^2_{H^0}},\ \ \
Q_H^2={k^2+k^{\prime2}\over k^2-k^{\prime2}},}
and we assumed $m_{H^0}\sim m_{A^0}\sim m_{H^+}$. The factor of 430
was first calculated in ref.
\ref\BBS{G. Beall, M. Bander and A. Soni, Phys. Rev. Lett. 48 (1982)
 848.},
and it is the product of three smaller  numbers: a factor of 2 since
two diagrams contribute, a factor of $4[\ln(m^2_{W_1}/m_c^2)-1]\sim28$
from loop integration and a factor of 7.6 due to the Lorentz
structure of the relevant matrix operator. The factor of 11600 arises
because $H^0$ contributes at tree level. The contribution from the
LRS diagrams could easily dominate $\epsK$. In order that it does
not give a too large contribution, we need
\eqn\epsLRbou{|\beta\sin(\delta_1-\delta_2)|\lsim10^{-5}.}
Note that in order that the real part of the same diagrams does not
give a too large $\Delta m_K$, we require $m_{W_2}\gsim1.7\ TeV$
and $m_H\gsim8.8\ TeV$. The bound $\beta\lsim1/430$ implies
\eqn\xibou{\xi\lsim2.2\times10^{-3}.}

The most important LRS contributions to $d_N$ arise from quark EDMs.
The LRS one-loop diagrams involve $W_L-W_R$ mixing and $W_R-d_R-
u_{iR}$ vertex, so all phases contribute, but $(\gamma+\delta_1)$ which
contributes proportionally to $m_c$ is the most important one
\nref\BeSo{G. Beall and A. Soni, Phys. Rev. Lett. 47 (1981) 552.}%
\nref\HMP{X.G. He, H.J. McKellar and S. Pakvasa,
 Phys. Rev. Lett. 61 (1988) 1267.}%
\refs{\BeSo-\HMP}:
\eqn\dNLRS{d_N\approx1.5\times10^{-21}\
\xi\sin(\gamma+\delta_1)\ e\ {\rm cm} .}
This could easily saturate the experimental bound (even with $\xi$ as
small as required by \xibou). In order not to violate the bound, we need
\eqn\dNLRbou{|\xi\sin(\gamma+\delta_1)|\lsim10^{-4}.}
There are also contributions to $d_N$ through the three gluon operator,
but these are about an order of magnitude smaller
\ref\CLY{D. Chang, C.S. Li and T.C. Yuan, Phys. Rev. D42 (1990) 867.}.

The LRS contribution to $\epe$, through {\it tree} diagrams involving
$W_L-W_R$ mixing, gives \refs{\Chan,\BFG,\HMP}
\eqn\epsLRS{|\epe|\approx276\xi|\sin(\gamma-\delta_2)
+\sin(\gamma-\delta_1)|.}
This could easily saturate the experimental bound (even with $\xi$ as
small as required by \xibou). In order not to violate the bound, we need
\eqn\epeLRbou{\xi|\sin(\gamma-\delta_1)
+\sin(\gamma-\delta_1)|\lsim10^{-5}.}

The effect of LRS on CP asymmetries in $B$ decays
\nref\EcGr{G. Ecker and W. Grimus, Z. Phys. C30 (1986) 293.}%
\nref\LoWy{D. London and D. Wyler, Phys. Lett. B232 (1989) 503.}%
\refs{\EcGr-\LoWy}\ is very small, because
LRS contributions to $B-\bar B$ mixing are small in magnitude. The reason
for that is as follows. One of the enhancement factors for the LRS
contribution to $K-\bar K$ mixing is the hadronic matrix element,
\eqn\enhLR{{\vev{K^0|\bar d_L s_R\bar d_R s_L|\bar K^0}\over
\vev{K^0|(\bar d_L \gamma^\mu s_L)^2|\bar K^0}}={3\over4}\left[
\left({m_K\over m_s+m_d}\right)^2+{1\over6}\right]\approx7.6.}
However, as $m_B\approx m_b$, there is no similar enhancement in the $B$
system. This implies that if LRS contributions to $K-\bar K$ mixing are
as large as the Standard Model ones, then the LRS contributions to
$B-\bar B$ mixing are $\O(0.1)$ of the Standard Model ones.

Finally, we mention that LRS effects on transverse lepton polarization
are negligible. This is due to the general result that vector
interactions alone cannot give any transverse lepton polarization
\Miriam. Other CP violating observables can be constructed with
potentially large effects. For this, an extra independent vector has to
be measured. Examples are four body semileptonic kaon decay \CFK, and $B$
and $D$ decays into vector mesons where the polarization of the vector
meson is measured \Garisto. The current bounds on the model parameters,
however, imply that these asymmetries are small \Garisto.

To summarize: even though all the phases in the LRS model with
spontaneously broken CP arise from a single phase $\eta$ in the
VEV $\vev{\Phi}$, it is difficult to relate their values unless
one makes additional assumptions. Thus, the three bounds that we
found, \epsLRbou, \dNLRbou\ and \epeLRbou, could all be saturated
simultaneously
\ref\LGN{J. Liu, C.Q. Geng and J.N. Ng, Phys. Rev. D39 (1989) 3473.}.
However, without (at least mild) fine-tuning, saturation of the $\epe$
bound would imply that the contribution to $d_N$ is two orders of
magnitude below the present experimental limit. If $k^\prime/k\lsim0.1$
and all phases are of the same order of magnitude, then the $\epsK$
bound is the strongest. If, furthermore, $k^\prime/k\ll m_s/m_c$,
then $\epe$ and $d_N$ are related \refs{\BFG,\HMP}\ through
\eqn\smallkp{|d_N|=3.6\times10^{-24}|\epe|\ e\ {\rm cm}.}
Finally, no interesting effects on $B-\bar B$ mixing are expected.

\newsec{Conclusions}

In this review, we studied various extensions of the Standard Model
and presented the new CP violating effects that are most likely to
occur in these extensions. When thinking of future measurements of
CP violating effects as a tool to discover New Physics, we should
distinguish between three classes of quantities:

(i) Observables with small theoretical uncertainties.
Here, New Physics effects can be unambiguously observed even if
they are comparable in magnitude or somewhat smaller than the
Standard Model contribution. The observables in this class are
mostly manifestations of interference between mixing and decay
in neutral meson decays.

\item{$\bullet$} CP asymmetries in specific $B$-meson decays such as
$B_d\to\psi K_S$, $B_d\to\pi\pi$ (with isospin analysis) and
$B_s\to\psi\phi$. These asymmetries are sensitive to extensions of
the quark sector, i.e. extra quarks in vector-like representations
or a fourth generation and to supersymmetric models where FCNC are
suppressed by alignment or by
heaviness of the first two squark generations.

\item{$\bullet$} The decay rate $K_L\to\pi\nu\nu$. This mode
is also sensitive to extensions of the quark sector.

(ii) Observables which are negligibly small (compared to the experimental
bound) in the Standard Model. New Physics effects will be signaled if
they are much larger than the Standard Model contributions.

\item{$\bullet$} Electric dipole moments. In particular, the electric
dipole moments of the neutron and the electron are likely to be close to
the experimental bound in supersymmetric models and in various
extensions of the Higgs sector.

\item{$\bullet$} Transverse lepton polarization. These quantities cannot
arise from vector interactions only and therefore are a sensitive probe
of extensions of the scalar sector.

\item{$\bullet$} CP violation in $D-\bar D$ mixing. These effects
will test the alignment mechanism to suppress FCNC in supersymmetry.
In addition, they might arise in extensions of the quark sector and
of the scalar sector.

(iii) Observables with large hadronic uncertainties that are not
negligibly small in the Standard Model. The observables in this
class are mostly related to CP violation in decay. Beyond their
usefulness to improve our understanding of the relevant hadronic
aspects, they may also exclude models of New Physics that predict
vanishingly small effects.

\item{$\bullet$} $\epe$.

\item{$\bullet$} Most CP asymmetries in charged $B$ decays.

The coming years hold great promise in clarifying the various puzzles
of CP violation. This may well turn to be the leading direction in
the search of New Physics.

\vskip 1 cm
\centerline{\bf Acknowledgements}
We thank Helen Quinn for useful comments on the manuscript.
Our understanding of CP violation has benefitted from collaborations
with Guy Blaylock, Claudio Dib, Lance Dixon, Michael Dine, Isi Dunietz,
Fred Gilman, Howard Haber, Haim Harari, Miriam Leurer, Zoltan Ligeti,
Zvi Lipkin, David London, Enrico Nardi, Helen Quinn, Uri Sarid,
Nati Seiberg, Abe Seiden, Yuri Shirman, Dennis Silverman, Art Snyder
and Mihir Worah. Y.G. is supported
by the Department of Energy under contract DE-AC03-76SF00515.
Y.N. is supported in part by the United States -- Israel Binational
Science Foundation (BSF), by the Israel Science Foundation,
and by the Minerva Foundation (Munich).

\listrefs
\end